\documentclass[11pt]{article}
\usepackage{verbatim,color}
\usepackage{amsthm,amsmath,amssymb,amsfonts}
\usepackage{graphicx}
\usepackage{epstopdf}   
\usepackage{bm}
\usepackage{natbib}
\bibpunct{(}{)}{;}{a}{}{,}
\usepackage[noline, ruled,boxed]{algorithm2e}
\usepackage[tight]{subfigure}
\usepackage[tableposition=top]{caption}
\usepackage{booktabs}
\usepackage[flushleft]{threeparttable}
\usepackage{arydshln}  
\usepackage{setspace}
\usepackage{afterpage}

\newtheorem{theorem}{Theorem}[section]
\newtheorem{proposition}[theorem]{Proposition}
\newtheorem{lemma}[theorem]{Lemma}

\def\Acal{\mathcal A}

\def\Hcal{\mathcal H}

\def\Rcal{\mathcal R}

\def\Xcal{\mathcal X}
\def\bE{\mathbb E}
\def\bI{\mathbb I}

\def\bR{\mathbb R}

\def\sign{{\rm sign}}

\def\argmin{\mathop{\rm argmin}}

\begin{document}

\def\spacingset#1{\renewcommand{\baselinestretch}%
{#1}\small\normalsize} \spacingset{1}


\title{\bf Augmented Outcome-weighted Learning for Optimal Treatment Regimes}
\author{Xin Zhou\\[5pt]
{Departments of Biostatistics and Epidemiology}\\
{Harvard T.H. Chan School of Public Health}\\
{Boston, Massachusetts 02115, U.S.A.}
\\[20pt]
Michael R. Kosorok\\[5pt]
{Department of Biostatistics}\\
{University of North Carolina at Chapel Hill}\\
{Chapel Hill, North Carolina 27599, U.S.A.}
\\[1pt]
}
\maketitle

\begin{abstract}
Precision medicine is of considerable interest in clinical, academic and regulatory parties.
The key to precision medicine is the optimal treatment regime.
Recently, \citet{Zhou2015:RWL} developed residual weighted learning (RWL) to construct the optimal regime that directly optimize
the clinical outcome.
However, this method involves computationally intensive non-convex optimization, which cannot guarantee a global solution.
Furthermore, this method does not possess fully semiparametrical efficiency.
In this article, we propose augmented outcome-weighted learning (AOL). The method is built on a doubly robust augmented inverse probability weighted estimator (AIPWE), and hence constructs semiparametrically efficient regimes.
Our proposed AOL is closely related to RWL. The weights are obtained from counterfactual residuals, where negative residuals are reflected to positive and accordingly their treatment assignments are switched to opposites. Convex loss functions are thus applied to guarantee a global solution and to reduce computations.
We show that AOL is universally consistent, \textit{i.e.}, the estimated regime of AOL converges the Bayes regime when the sample size approaches infinity, without knowing any specifics of the distribution of the data.
We also propose variable selection methods for linear and nonlinear regimes, respectively, to further improve performance.
The performance of the proposed AOL methods is illustrated in simulation studies and in an analysis of
the Nefazodone-CBASP clinical trial data.
\end{abstract}

\noindent%

\section{Introduction}
\label{sec:introduction}
Most medical treatments are designed for the ``average patient''. Such a ``one-size-fits-all'' approach is successful for some patients but not always for others. Precision medicine, also known as personalized medicine, is an innovative approach to disease prevention and treatment that take into account individual variability in clinical information, genes, environments and lifestyles. Currently, precision medicine is of considerable interest in clinical, academic, and regulatory parties. There are already several FDA-approved treatments that are tailored to specific characteristics of individuals. For example, ceritinib, a recently FDA approved drug for the treatment of lung cancer, is highly active in patients with advanced, ALK-rearranged non-small-cell lung cancer \citep{Shaw2014:Ceritinib}.

The key to precision medicine is the optimal treatment regime. Let $\bm{X}=(X_1,\cdots,X_p)^T\in\mathcal{X}$ be a patient's clinical covariates, $A\in\mathcal{A}=\{+1,-1\}$ be the treatment assignment, and $R$ be the observed clinical outcome. Assume without loss of generality that larger values of $R$ are more desirable. A treatment regime $d$ is a function from $\mathcal{X}$ to $\mathcal{A}$. An optimal treatment regime is a regime that maximizes the outcome under this regime.
Assuming that the data generating mechanism is known, the optimal treatment regime is related to the contrast
\begin{equation*}
\delta(\bm{x}) = \mu_{+1}(\bm{x})-\mu_{-1}(\bm{x}),
\end{equation*}
where $\mu_{+1}(\bm{x})=\mathbb{E}(R|\bm{X}=\bm{x},A=+1)$ and $\mu_{-1}(\bm{x})=\mathbb{E}(R|\bm{X}=\bm{x},A=-1)$.
The Bayes optimal regime is $d^{\ast}(\bm{x}) = 1$ if $\delta(\bm{x})>0$ and $-1$ otherwise.

Most of published optimal treatment strategies estimate the contrast $\delta(\bm{x})$ by modelling either the conditional mean outcomes or contrast directly based on data from randomized clinical trials or observational studies (see \citet{Moodie2014:Qlearning, Murphy:OptimalDTR2003, Robins2004:SNM, Taylor2015:QRF} and references therein). They obtain treatment regimes indirectly by inverting the regression estimates.  They are regression-based approaches for treatment regimes. For instance, \citet{Qian:ITR2011} proposed a two-step procedure
that first estimates a conditional mean for the outcome and then determines the treatment regime
by comparing conditional mean outcomes across various treatments. The success of these regression-based approaches depends
on the correct specification of models and on the high precision of the model estimates. However, in practice, the heterogeneity in population makes the regression model estimate complicated.

Alternatively, \citet{Zhao:OWL2012} proposed a classification-based approach, called outcome weighted learning (OWL), to
utilize the weighted support vector machines \citep{Vapnik:svm95} to estimate the optimal treatment regime directly. \citet{Zhang2012:ClassificationITR} also proposed a general framework to make use of classification methods to the optimal treatment regime problem.

\begin{figure}
\centering
\includegraphics[scale=0.6]{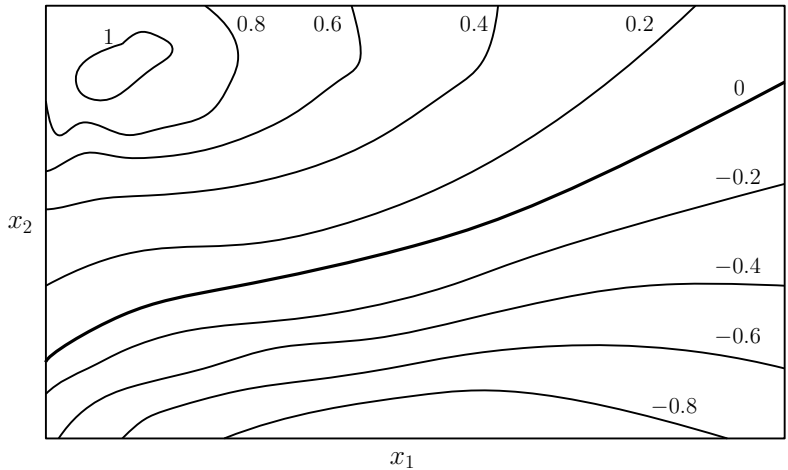}
\caption{Example of contour of $\delta(\bm{x})$ on $\bm{x}=(x_1,x_2)^T$. The decision boundary is $\delta(\bm{x})=0$. The optimal treatment is $1$ if $\delta(\bm{x})>0$, and $-1$ otherwise. The decision boundary can be approximated by a linear function, although the contrast function $\delta(\bm{x})$ is not linear on $\bm{x}$. }\label{fig:example}
\end{figure}

Indeed, the classification-based approaches follow Vapnik's main principle \citep{Vapnik:svm95}: ``When solving a given problem, try to avoid solving a more general problem as an intermediate step.'' As in Figure \ref{fig:example}, the aim of optimal treatment regimes is to estimate the form of the decision boundary $\delta(\bm{x})=0$. Regression-based approaches find the decision boundary by solving a general problem that estimates $\delta(\bm{x})$ for any $\bm{x}\in\mathcal{X}$. For the optimal treatment regime, it is sufficient to find an accurate estimate of $\delta(\bm{x})$ only near the zeros of $\delta(\bm{x})$. In general, finding the optimal regime is an easier problem than regression function estimation.
Classification-based approaches, which seek the decision boundary directly, provide a flexible framework from a different perspective.

Recently, \citet{Zhou2015:RWL} proposed residual weighted learning (RWL), which uses the residual from a regression fit of outcome as the pseudo-outcome, to improve finite sample performance of OWL. However, this method, involving a non-convex loss function, presents numerous challenges in computations, which hinders its practical use. For non-convex optimization, a global solution is not guaranteed, and the computation is generally intensive.
\citet{Athey2017:policy} also pointed out that RWL does not possess fully semiparametrical efficiency.
In this article, we propose augmented outcome-weighted learning (AOL). The method is built on a doubly robust augmented inverse probability weighted estimator (AIPWE), and hence constructs semiparametrically efficient regimes.
Although this article focuses on randomized clinical trials, the double robustness is particularly useful for observational studies.
Our proposed AOL is closely related to RWL. The weights are obtained from counterfactual residuals, where negative residuals are reflected to positive and accordingly their treatment assignments are switched to opposites. Convex loss functions are thus applied to reduce computations.
AOL inherits almost all desirable properties of RWL. Similar with RWL, AOL is also universally consistent, \textit{i.e.}, the estimated regime of AOL converges the Bayes regime when the sample size approaches infinity, without knowing any specifics of the distribution of the data. The finite sample performance of AOL is demonstrated in numerical simulations.

The remainder of the article is organized as follows. In Section 2.1, we review outcome weighted learning and residual weighted learning. In Section 2.2 and 2.3, we propose augmented outcome-weighted learning. We discover the connection between augmented outcome-weighted learning and residual weighted learning in Section 2.4.
We establish universal consistency for the proposed AOL in Section 2.5. The variable selection techniques for AOL are discussed in Section 2.6. We present simulation studies to evaluate finite sample performance of the proposed methods in Section 3. The method is then illustrated on the Nefazodone-CBASP clinical trial in Section 4. We conclude the article with a discussion in Section 5. All the technical proofs are provided in Appendix.


\section{Method}
\label{sec:method}
\subsection{Review of outcome weighted learning and residual weighted learning}
In this article, random variables are denoted by uppercase letters, while their realizations are denoted by lowercase letters.
Consider a two-arm randomized trial.
Let $\pi(a,\bm{x}):=P(A=a|\bm{X}=\bm{x})$ be the probability of being assigned treatment $a$ for patients with clinical covariates $\bm{x}$. It is predefined in the trial design. We assume $\pi(a,\bm{x})>0$ for all $a\in\mathcal{A}$ and $\bm{x}\in\mathcal{X}$.

We use the potential outcomes framework \citep{Rubin1974:PotentialOutcome} to precisely define the optimal treatment regime. Let $R^{\ast}(+1)$ and $R^{\ast}(-1)$ denote the potential outcomes that would be observed had a subject received treatment $+1$ or $-1$. There are two assumptions in the framework. The actually observed outcomes and potential outcomes are connected by the consistency assumption, \textit{i.e.}, $R=R^{\ast}(A)$. We further assume that conditional on covariates $\bm{X}$, the potential outcomes $\{(R^{\ast}(+1), R^{\ast}(-1)\}$ are independent of $A$, the treatment that has been actually received. This is the assumption of no unmeasured confounders (NUC). This assumption is automatically hold in a randomized clinical trial.

For an arbitrary treatment regime $d$, we can thus define its potential outcome $R^{\ast}(d(\bm{X})) =R^{\ast}(+1)\bI({d(\bm{X})=+1})+R^{\ast}(-1)\bI({d(\bm{X})=-1})$, where $\mathbb{I}(\cdot)$ is the indicator function.
It would be the observed outcome if a subject from the population were to be assigned treatment according
to regime $d$.
The expected potential outcome under any regime $d$, defined as $\mathcal{V}(d) = \bE(R^{\ast}(d))$, is called
the value function associated with regime $d$. Thus, an optimal regime $d^{\ast}$ is a regime that maximizes $\mathcal{V}(d)$.
Let $m(\bm{x},d)=\mu_{+1}(\bm{x})\bI({d(\bm{x})=+1})+\mu_{-1}(\bm{x})\bI({d(\bm{x})=-1})$.
Under the consistency and NUC assumptions, it is straightforward to show that
\begin{equation} \label{eq:value}
\mathcal{V}(d) = \bE\Big(m(\bm{X},d))\Big) = \mathbb{E}\left(\frac{R}{\pi(A,\bm{X})}\mathbb{I}\Big(A=d(\bm{X})\Big)\right).
\end{equation}
Thus finding $d^{\ast}$ is equivalent to the following minimization problem:
\begin{equation}\label{qieq:weightederror}
d^{\ast}\in \arg\min_{d}\,\mathbb{E}\left(\frac{R}{\pi(A,\bm{X})}\mathbb{I}\Big(A\neq d(\bm{X})\Big)\right).
\end{equation}
\citet{Zhao:OWL2012} viewed this as a weighted classification problem, and proposed outcome weighted learning (OWL) to apply statistical learning techniques to optimal treatment regimes.
However, as discussed in \citet{Zhou2015:RWL}, this method is not perfect. Firstly, the estimated regime of OWL is affected by a simple shift of the outcome $R$. Hence estimates from OWL are unstable especially when the sample size is small.
Secondly, since OWL needs the outcome to be nonnegative to gain computational efficiency from convex programming, OWL works similarly as weighted classification to reduce the difference between the estimated and true treatment assignments. Thus the regime by OWL tends to retain the treatments that subjects actually received. This behavior is not ideal for data from a randomized clinical trial, since treatments are actually randomly assigned to patients.

To alleviate these problems, \citet{Zhou2015:RWL} proposed residual weighted learning (RWL), in which the misclassification errors are weighted by residuals of the outcome $R$ from a regression fit on clinical covariates $\bm{X}$. The residuals are calculated as
\begin{equation*} 
{R}_g = R - {g}(\bm{X}).
\end{equation*}
\citet{Zhou2015:RWL} used ${g}_1(\bm{X})=\mathbb{E}({R \over 2\pi(A,\bm{X})}|\bm{X})$ as a choice of $g(\bm{X})$.
Unlike OWL in (\ref{qieq:weightederror}), RWL targets the following optimization problem,
\begin{equation*} 
d^*\in \arg\min_{d}\,\mathbb{E}\left(\frac{{R}_g}{\pi(A,\bm{X})}\mathbb{I}\Big(A\neq d(\bm{X})\Big)\right).
\end{equation*}

Suppose that the realization data $\{(\bm{x}_i, a_i, r_i): i=1,\cdots,n\}$  are collected independently. For any decision function $f(\bm{x})$, let $d_f(\bm{x})=\textrm{sign}\big(f(\bm{x})\big)$ be the associated regime. RWL aims to minimize the following regularized empirical risk,
\begin{equation} \label{qieq:rwloptimization}
\frac{1}{n}\sum_{i=1}^{n}\frac{{r}_{g,i}}{\pi(a_i,\bm{x}_i)}T\Big(a_if(\bm{x}_i)\Big)+\lambda||f||^2,
\end{equation}
where ${r}_{g,i}={r}_{i}-g(\bm{x}_i)$, $T(\cdot)$ is a continuous surrogate loss function, $||f||$ is some norm for $f$, and $\lambda$ is a tuning parameter.

Since some residuals are negative, convex surrogate loss functions are not appropriate in (\ref{qieq:rwloptimization}). \citet{Zhou2015:RWL} considered a non-convex loss, the smoothed ramp loss function.
However, the non-convexity presents significant challenges for solving the optimization problem (\ref{qieq:rwloptimization}). Unlike convex functions, non-convex functions may possess local optima that are not global optima, and most of efficient optimization algorithms, such as gradient descent and coordinate descent, are only guaranteed to converge to a local optimum. The theoretical properties of RWL establish on the global optimum. Although \citet{Zhou2015:RWL} applied a difference of convex (d.c.) algorithm to address the non-convex optimization problem by solving a sequence of convex subproblems to increase the likelihood of reaching a global minimum, the global optimization is not guaranteed \citep{Sriperumbudur2009:CCCP}. The d.c. algorithm is still computationally intensive. In addition, RWL may connect with AIPWE as discussed in \citet{Zhou2015:RWL}, but it does not have fully semiparametrical efficiency \citep{Athey2017:policy}.

\subsection{Augmented outcome-weighted learning (AOL)} \label{sec:aol}
Let us come back to equation \eqref{eq:value}. The first equality is the foundation of regression-based approaches, while the second inspired outcome weighted learning \citep{Zhao:OWL2012}. \cite{Zhang2012:ClassificationITR} combined these two perspectives through a doubly robust augmented inverse probability weighted estimator \citep[AIPWE]{Bang2005:DRE} of the value function.

Recall that $\mu_{+1}(\bm{x})=\mathbb{E}(R|\bm{X}=\bm{x},A=+1)$, $\mu_{-1}(\bm{x})=\mathbb{E}(R|\bm{X}=\bm{x},A=-1)$, and $m(\bm{x},d)=\mu_{+1}(\bm{x})\bI({d(\bm{x})=+1})+\mu_{-1}(\bm{x})\bI({d(\bm{x})=-1})$.
Following \citet{Zhang:RobustITR2012}, we start from the doubly robust AIPWE:
\begin{equation*}
\textrm{AIPWE}(d)=\frac{1}{n}\sum_{i=1}^n\left(\frac{r_i-\hat{m}(\bm{x}_i,d)}{\pi(a_i,\bm{x}_i)}\bI\big(a_i=d(\bm{x}_i)\big)+\hat{m}(\bm{x}_i,d)\right),
\end{equation*}
where $\hat{m}(\bm{x},d)$ is an estimator of ${m}(\bm{x},d)$, which is an estimator of
\begin{equation*} \label{eq:AIPWE}
\mathcal{V}(d) = \bE\left(\frac{R-{m}(\bm{X},d)}{\pi(A,\bm{X})}\bI(A=d(\bm{X}))+{m}(\bm{X},d)\right).
\end{equation*}
For an observational study, we are also required to estimate $\pi(a,\bm{x})$ by the data. $\textrm{AIPWE}(d)$ is a consistent estimator of $\mathcal{V}(d)$ if either $\hat\pi(a,\bm{x})$ or $\hat{m}(\bm{x},d)$ is correctly specified.
This is the so-called double robustness.
In a randomized clinical trial $\pi(a,\bm{x})$ is known, hence even if $\hat{m}(\bm{x},d)$ is inconsistent, $\textrm{AIPWE}(d)$ is still consistent.

Noting that
\begin{equation*}
\frac{R-{m}(\bm{X},d)}{\pi(A,\bm{X})}\bI\big(A=d(\bm{X})\big)+{m}(\bm{X},d) = \frac{R-\tilde{g}(\bm{X})}{\pi(A,\bm{X})}\bI\big(A=d(\bm{X})\big)+\mu_{-A}(\bm{X}),
\end{equation*}
where
\begin{equation} \label{eq:tildeg}
\tilde{g}(\bm{x}):=\pi(-1,\bm{x})\mu_{+1}(\bm{x})+\pi(+1,\bm{x})\mu_{-1}(\bm{x}),
\end{equation}
maximizing $\textrm{AIPWE}(d)$ is asymptotic to the following minimization problem
\begin{equation} \label{eq:AOL0}
\arg\min_{d} \quad \bE\left(\frac{R-\tilde{g}(\bm{X})}{\pi(A,\bm{X})}\bI\big(A\neq d(\bm{X})\big)\right).
\end{equation}

Let $\tilde{R} = R - \tilde{g}(\bm{X})$.
As explained later in Section \ref{sec:connection}, $\tilde{R}$ is a form of residuals.
At this point, we may apply a similar non-convex surrogate loss in the regularization framework as RWL in \eqref{qieq:rwloptimization}. However, it still suffers from local optimization and intensive computation.

To seek the optimal regime, we apply a finding in \citet{Liu2014:ABSOWL} to take advantage of efficient convex optimization. Note that
\begin{equation*}
\mathbb{E}\left({|\tilde{R}|\over \pi(A,\bm{X})}\mathbb{I}\Big(A\cdot\textrm{sign}(\tilde{R})\neq d(\bm{X})\Big)\right) = \mathbb{E}\left({\tilde{R}\over \pi(A,\bm{X})}\mathbb{I}\Big(A\neq d(\bm{X})\Big)\right) + \mathbb{E}\left({\tilde{R}^-\over \pi(A,\bm{X})}\right),
\end{equation*}
where $\tilde{R}^- = \max(-\tilde{R}, 0)$. Therefore finding $d^*$ in (\ref{eq:AOL0}) is equivalent to the following optimization problem,
\begin{equation*} 
d^*\in \arg\min_{d}\,\mathbb{E}\left(\frac{|\tilde{R}|}{\pi(A,\bm{X})}\mathbb{I}\Big(A\cdot\textrm{sign}(\tilde{R})\neq d(\bm{X})\Big)\right),
\end{equation*}
where negative weights are reflected to positive, and accordingly their treatment assignments are switched to opposites.

Similar with OWL and RWL, we seek the decision function $f$ by minimizing a regularized surrogate risk,
\begin{equation}\label{qieq:empiriclAOL}
\frac{1}{n}\sum_{i=1}^{n}\frac{|\tilde{r}_{i}|}{\pi(a_i,\bm{x}_i)}\phi\Big(a_i\cdot\textrm{sign}(\tilde{r}_{i})f(\bm{x}_i)\Big)+\frac{\lambda}{2}||f||^2,
\end{equation}
where $\phi(\cdot)$ is a continuous surrogate loss function, $||f||$ is some norm for $f$, and $\lambda$ is a tuning parameter controlling the trade-off between the empirical risk and the complexity of the decision function $f$.
This method is called augmented outcome-weighted learning (AOL) in this article, since the weights are derived from augmented outcomes.

As the weights $\frac{|\tilde{r}_i|}{\pi(a_i,\bm{x}_i)}$ are all nonnegative, convex surrogate can be employed for efficient computation.
In this article, we apply the Huberized hinge loss function \citep{Wang2008:hhsvm},
\begin{equation} \label{eq:hhloss}
\phi(u) = \left\{
\begin{array}{ll}
0 & {\rm if}\: u\geq1, \\
\frac{1}{4}(1-u)^2 & {\rm if}\: -1\leq u<1, \\
-u & {\rm if}\: u<-1.
\end{array}
\right.
\end{equation}
Other convex loss functions, such as the hinge loss, can be also applied in AOL. Although the Huberized hinge loss has a similar shape with the hinge loss, the Huberized hinge loss is smooth everywhere. Hence it has computational advantages in optimization.

\subsection{Implementation of AOL}
We derive an algorithm for the linear AOL in Section~\ref{sec:linearaol}, and then generalize it to the case of nonlinear learning through kernel mapping in Section~\ref{sec:nonlinearaol}. Both algorithms solve convex optimization problems, and global solutions are guaranteed.
\subsubsection{Linear Decision Rule for AOL} \label{sec:linearaol}
Consider a linear decision function $f(\bm{x})=\bm{w}^T\bm{x}+b$. The associated regime $d_f$ will assign a subject with clinical covariates $\bm{x}$ into treatment 1 if $\bm{w}^T\bm{x}+b>0$ and $-1$ otherwise. In (\ref{qieq:empiriclAOL}), we define $||f||$ as the Euclidean norm of $\bm{w}$. Then the minimization problem (\ref{qieq:empiriclAOL}) can be rewritten as
\begin{equation} \label{qieq:linearAOL}
\min_{\bm{w},b}\quad\frac{1}{n}\sum_{i=1}^{n}\frac{|\tilde{r}_{i}|}{\pi(a_i,\bm{x}_i)}\phi\Big(a_i\cdot\textrm{sign}(\tilde{r}_{i})\big(\bm{w}^T\bm{x}_i+b\big)\Big)+ \frac{\lambda}{2}\bm{w}^T\bm{w}.
\end{equation}
There are many efficient numerical methods for solving this smooth unconstrained convex optimization problem. One example is the limited-memory Broyden-Fletcher-Goldfarb-Shanno (L-BFGS) algorithm \citep{Nocedal1980:LBFGS}, a quasi-Newton method that approximates the Broyden-Fletcher-Goldfarb-Shanno (BFGS) algorithm using a limited amount of computer memory. When we obtain the solution $(\hat{\bm{w}},\hat{b})$, the decision function is $\hat{f}(\bm{x})=\hat{\bm{w}}^T\bm{x}+\hat{b}$.

\subsubsection{Nonlinear Decision rule for AOL} \label{sec:nonlinearaol}
The nonlinear decision function $f(\bm{x})$ can be represented by $h(\bm{x})+b$ with $h(\bm{x})\in \mathcal{H}_K$ and $b\in\bR$, where $\mathcal{H}_K$ is a reproducing kernel Hilbert space (RKHS) associated with a Mercer kernel function $K$. The kernel function $K(\cdot,\cdot)$ is a positive definite function mapping from $\mathcal{X}\times\mathcal{X}$ to $\mathbb{R}$. The norm in $\mathcal{H}_K$, denoted by $||\cdot||_K$, is induced by the following inner product,
\begin{equation*}
<f,g>_K=\sum_{i=1}^{n}\sum_{j=1}^{m}\alpha_i\beta_jK(\bm{x}_i,\bm{x}_j),
\end{equation*}
for $f(\cdot)=\sum_{i=1}^n\alpha_iK(\cdot,\bm{x}_i)$ and $g(\cdot)=\sum_{j=1}^m\beta_jK(\cdot,\bm{x}_j)$.
The most widely used nonlinear kernel in practice is the Gaussian Radial Basis Function (RBF) kernel, that is,
\begin{equation*}
K_{\sigma}(\bm{x},\bm{z})=\exp\Big(-\sigma^2||\bm{x}-\bm{z}||^2\Big),
\end{equation*}
where $\sigma>0$ is a free parameter whose inverse $1/\sigma$ is called the width of $K_{\sigma}$.

Then minimizing (\ref{qieq:empiriclAOL}) can be rewritten as
\begin{equation}
\min_{h,b} \quad
\frac{1}{n}\sum_{i=1}^{n}\frac{|\tilde{r}_{g,i}|}{\pi(a_i,\bm{x}_i)}\phi\Big(a_i\cdot\textrm{sign}(\tilde{r}_{i})\big(h(\bm{x}_i)+b\big)\Big)
+\frac{\lambda}{2}||h||_K^2.
\label{qieq:nonlinearAOL0}
\end{equation}
Due to the representer theorem  \citep{Kimeldorf:Representer}, the nonlinear problem can be reduced to finding finite-dimensional coefficients $v_i$, and $h(\bm{x})$ can be represented as $\sum_{j=1}^nv_jK(\bm{x},\bm{x}_j)$. So the problem~(\ref{qieq:nonlinearAOL0}) is changed to
\begin{equation}
\min_{\bm{v},b} \quad \frac{1}{n}\sum_{i=1}^{n}\frac{|\tilde{r}_{i}|}{\pi(a_i,\bm{x}_i)}\phi\Big(a_i\cdot\textrm{sign}(\tilde{r}_{i})\big(\sum_{j=1}^nv_jK(\bm{x}_i,\bm{x}_j)+b\big)\Big)+
\frac{\lambda}{2}\sum_{i,j=1}^nv_iv_jK(\bm{x}_i,\bm{x}_j).
\label{qieq:nonlinearAOL}
\end{equation}
Again, it is a smooth unconstrained convex optimization problem. We apply L-BFGS algorithm to solve (\ref{qieq:nonlinearAOL}). When we obtain the solution $(\hat{\bm{v}},\hat{b})$, the decision function is $\hat{f}(\bm{x})=\sum_{j=1}^n\hat{v}_jK(\bm{x},\bm{x}_j)+\hat{b}$.

\subsection{Connection to residual weighted learning} \label{sec:connection}
Note that $\tilde{g}(\bm{x})$ in \eqref{eq:tildeg} is a weighted average of $\mu_{+1}(\bm{x})$ and $\mu_{-1}(\bm{x})$. Hence $\tilde{R}=R-\tilde{g}(\bm{X})$ is a form of residuals. The use of residuals in optimal treatment regimes is justified in \citet{Zhou2015:RWL} as follows, for any measurable function $g$,
\begin{equation*}
\mathbb{E}\left({R-g(\bm{X})\over \pi(A,\bm{X})}\mathbb{I}\Big(A\neq d(\bm{X})\Big)\right) = 
\mathbb{E}\left({R\over \pi(A,\bm{X})}-g(\bm{X})\right)-\mathcal{V}(d).
\end{equation*}
For residual weighted learning in \citet{Zhou2015:RWL}, the corresponding $g(\cdot)$ is
\begin{equation} \label{eq:weightrwl}
{g}_1(\bm{x}) = \mathbb{E}\left(\frac{R}{2\pi(A,X)}\big|\bm{X}=\bm{x}\right) = \frac{1}{2}\mu_{+1}(\bm{x})+\frac{1}{2}\mu_{-1}(\bm{x}).
\end{equation}
Similarly, \citet{Liu2014:ABSOWL} applied unweighted regression to calculate residuals, where the corresponding $g(\cdot)$ is
\begin{equation} \label{eq:weightreg}
{g}_2(\bm{x}) = \mathbb{E}\left(R|\bm{X}=\bm{x}\right) = \pi(+1,\bm{x})\mu_{+1}(\bm{x})+\pi(-1,\bm{x})\mu_{-1}(\bm{x}).
\end{equation}
It is interesting to understand the implication of $\tilde{g}(\bm{x})$ in \eqref{eq:tildeg}. Under the consistency and NUC assumptions, we can check that
\begin{equation*}
\mathbb{E}(R^{\ast}(-A)|\bm{X}=\bm{x}) =  \pi(-1,\bm{x})\mu_{+1}(\bm{x})+\pi(+1,\bm{x})\mu_{-1}(\bm{x}) = \tilde{g}(\bm{x}).
\end{equation*}
$\tilde{g}(\bm{x})$ is the expected outcome for subjects with covariate $\bm{x}$ had they received the opposite treatments to the ones that they have actually received. $\tilde{g}(\bm{x})$ is counterfactual, and cannot be observed. It can be estimated by $\hat{\tilde{g}}(\bm{x})=\pi(-1,\bm{x})\hat\mu_{+1}(\bm{x})+\pi(+1,\bm{x})\hat\mu_{-1}(\bm{x})$, where $\hat\mu_{+1}(\bm{x})$ and $\hat\mu_{-1}(\bm{x})$ are estimates of $\mu_{+1}(\bm{x})$ and $\mu_{-1}(\bm{x})$, respectively. Noting that
\begin{equation} \label{eq:weightaol}
\tilde{g}(\bm{x})=\mathbb{E}\left(\frac{\pi(-A,\bm{X})}{\pi(A,\bm{X})}R\big|\bm{X}=\bm{x}\right) = \pi(-1,\bm{x})\mu_{+1}(\bm{x})+\pi(+1,\bm{x})\mu_{-1}(\bm{x}),
\end{equation}
$\tilde{g}(\bm{x})$ also can be estimated by weighted regression directly, where weights are $\frac{\pi(-A,\bm{x})}{\pi(A,\bm{x})}$.
In a randomized clinical trial with usual equal allocation ratio 1:1, $g_1(\bm{x})$, $g_2(\bm{x})$ and $\tilde{g}(\bm{x})$ coincide. If the allocation ratio is unequal, they are different.
Compared with the regression weights of $g_1(\bm{x})$ in \eqref{eq:weightrwl} and of $g_2(\bm{x})$ in \eqref{eq:weightreg}, $\tilde{g}(\bm{x})$ in \eqref{eq:weightaol} utilizes a more extreme set of weights. For example, in a randomized clinical trial with the allocation ratio $3:1$, \textit{i.e.,} the number of subjects in arm $+1$ is three times as that in arm $-1$, the weights in \eqref{eq:weightreg} for two arms are both 1 (unweighted), the weights in \eqref{eq:weightrwl} are $2/3$ and 2, and the weights in \eqref{eq:weightaol} are $1/3$ and 3.

Our proposed AOL is closely related to RWL, as we just discussed that AOL uses counterfactual residuals.
AOL possesses almost all desirable properties of RWL. First, by using residuals, AOL stabilizes the variability introduced from the original outcome. Second, to minimize the empirical risk in (\ref{qieq:empiriclAOL}), for subjects with positive residuals, AOL tends to recommend the same treatment assignments that subjects have actually received; for subjects with negative residuals, AOL is apt to give the opposite treatment assignments to what they have received.
Third, AOL is location-scale invariant with respect to the original outcomes. Specifically, the estimated regime from AOL is invariant to a shift of the outcome; it is invariant to a scaling of the outcome with a positive number; the regime from AOL that maximizes the outcome is opposite to the one that minimizes the outcome. These are intuitively sensible.
The only nice property of RWL that is not inherited by AOL is the robustness to outliers because of the unbounded convex loss in AOL. However, we may apply an appropriate method or model estimating residuals to reduce the probability of outliers.

\subsection{Theoretical properties} \label{sec:theory}
In this section, we establish theoretical properties for AOL.
Recall that for any treatment regime $d:\mathcal{X}\rightarrow\mathcal{A}$, the value function is defined as
\begin{equation*}
\mathcal{V}(d) = \bE\left(\frac{R}{\pi(A,\bm{X})}\mathbb{I}\Big(A=d(\bm{X})\Big)\right).
\end{equation*}
Similarly, we define the risk function of a treatment regime $d$ as
\begin{equation*}
\mathcal{R}(d) = \bE\left(\frac{R}{\pi(A,\bm{X})}\mathbb{I}\Big(A\neq d(\bm{X})\Big)\right).
\end{equation*}
The regime that minimizes the risk is the Bayes regime $d^{\ast}=\arg\min_d\mathcal{R}(d)$, and the corresponding risk $\mathcal{R}^{\ast}=\mathcal{R}(d^{\ast})$ is the Bayes risk.
Recall that the Bayes regime is $d^{\ast}(\bm{x}) = 1$ if $\delta(\bm{x})>0$ and $-1$ otherwise.

Let $\phi: \bR\mapsto\bR_+$, where $\bR_+=[0,+\infty)$, be a convex function. In this section, we investigate a general result, and do not limit $\phi$ as the Huberized hinge loss. A few popular convex surrogate examples are listed as follows:
\begin{itemize}
\item Hinge loss: $\phi(u)=(1-u)_+$, where $(v)_+=\max(0,v)$,
\item Squared hinge loss: $\phi(u)=[(1-u)_+]^2$,
\item Least squares loss: $\phi(u)=(1-u)^2$,
\item Huberized hinge loss as shown in (\ref{eq:hhloss}),
\item Logistic loss: $\phi(u)=\log(1+\exp(-u))$,
\item Distance weighted discrimination (DWD) loss:
\begin{equation*}
\phi(u) = \left\{
\begin{array}{ll}
\frac{1}{u} & {\rm if}\: u\geq1, \\
2-u & {\rm if}\: u<1,
\end{array}
\right.
\end{equation*}
\item Exponential loss: $\phi(u)=\exp(-u)$.
\end{itemize}
The hinge loss and squared hinge loss are widely used in support vector machines \citep{Vapnik:svm95}. The least squares loss is applied to regularization networks \citep{Evgeniou:rnsvm}. The loss function in the logistic regression is just the logistic loss. The DWD loss is the loss function in the distance-weighted discrimination \citep{Marron2007:DWD}.
The exponential loss is used in AdaBoost \citep{Freund1997:AdaBoost}.

For any measurable function $g: \Xcal\mapsto \bR$, recall that ${R}_g = R - g(\bm{x})$. In this section, we do not require $g$ to be a regression fit of $R$, and $g$ can be any arbitrary function.
For a decision function $f: \Xcal\mapsto \bR$, we proceed to define a surrogate $\phi$-risk function:
\begin{equation} \label{eq:phirisk}
\Rcal_{\phi,g}(f)=\bE\left(\frac{|{R}_g|}{\pi(A,\bm{X})}{\phi\Big(A\cdot\sign({R}_g)f(\bm{X})\Big)}\right).
\end{equation}
Similarly, the minimal $\phi$-risk as $\Rcal_{\phi,g}^{\ast}=\inf_f\Rcal_{\phi,g}(f)$ and $f_{\phi,g}^{\ast}=\arg\min_f\Rcal_{\phi,g}(f)$.

The performance of the associated regime $d_f=\textrm{sign}(f)$ is measured by the excess risk $\Delta\Rcal(f)=\mathcal{R}(d_f)-\mathcal{R}^{\ast}$. Similarly, we define the excess $\phi$-risk as $\Delta\Rcal_{\phi,g}(f)=\mathcal{R}_{\phi,g}(f)-\mathcal{R}_{\phi,g}^{\ast}$.

Suppose that a sample $\mathcal{D}_n=\{X_i, A_i, R_i\}_{i=1}^n$ is independently drawn from a probability measure $P$ on $\Xcal\times\Acal\times\bR$, where $\Xcal\subset\bR^p$ is compact.
Let $f_{D_n,\lambda_n}\in\Hcal_K+\{1\}$, \textit{i.e.} $f_{D_n,\lambda_n}=h_{D_n,\lambda_n}+b_{D_n,\lambda_n}$, where $h_{D_n,\lambda_n}\in\Hcal_K$ and $b_{D_n,\lambda_n}\in\bR$, be a global minimizer of the following optimization problem:
\begin{equation} \label{eq:AOLHk}
\min_{f=h+b\in\Hcal_K+\{1\}} \quad \frac{1}{n}\sum_{i=1}^n\frac{|{R}_{g,i}|}{\pi(A_i,\bm{X}_i)}{\phi\Big(A_i\cdot\sign({R}_{g,i})f(\bm{X}_i)\Big)}+\frac{\lambda_n}{2}||h||_K^2,
\end{equation}
where ${R}_{g,i}=R_i-g(\bm{X}_i)$. Here we suppress $\phi$ and $g$ from the notations of $f_{D_n,\lambda_n}$, $h_{D_n,\lambda_n}$ and $b_{D_n,\lambda_n}$.

The purpose of the theoretical analysis is to investigate universal consistency of the associated regime of $f_{D_n,\lambda_n}$. The concept of universal consistency is given in \citet{Zhou2015:RWL}. A universally consistent regime method eventually learns the Bayes regime without knowing any specifics of the distribution of the data when the sample size approaches infinity. Mathematically, a regime $d$ is universally consistent when $\lim_{n\rightarrow\infty} \mathcal{R}(d)=\mathcal{R}^{\ast}$ in probability.

\subsubsection{Fisher consistency}
The first question is whether the loss function used is Fisher consistent. The concept of Fisher consistency is brought from pattern classification \citep{Lin:svmbayes}. For optimal treatment regimes, a loss function is Fisher consistent if the loss function alone can be used to identify the Bayes regime when the sample size approaches infinity, \textit{i.e.}, $\mathcal{R}(\textrm{sign}(f_{\phi,g}^{\ast})) = \mathcal{R}(d^{\ast})$.
We define
\begin{eqnarray} \label{eq:eta}
\eta_{1}(\bm{x}) & = & \mathbb{E}({R}_g^+|\bm{X}=\bm{x},A=+1) + \mathbb{E}({R}_g^-|\bm{X}=\bm{x},A=-1), \nonumber \\
\eta_{2}(\bm{x}) & = & \mathbb{E}({R}_g^+|\bm{X}=\bm{x},A=-1) + \mathbb{E}({R}_g^-|\bm{X}=\bm{x},A=+1),
\end{eqnarray}
where ${R}_g^+=\max({R}_g,0)$ and ${R}_g^-=\max(-{R}_g,0)$. We suppress the dependence on $g$ from the notations.
Note that
\begin{equation*}
\eta_{1}(\bm{x})-\eta_{2}(\bm{x})=\bE(R|\bm{X}=\bm{x},A=+1)-\bE(R|\bm{X}=\bm{x},A=-1)=\mu_{+1}(\bm{x})-\mu_{-1}(\bm{x}).
\end{equation*}
The sign of $\eta_{1}(\bm{x})-\eta_{2}(\bm{x})$ is just the Bayes regime on $\bm{x}$. After some simple algebras, the $\phi$-risk in (\ref{eq:phirisk}) can be shown as
\begin{equation*}
\Rcal_{\phi,g}(f)=\bE\Big(\eta_{1}(\bm{X})\phi\big(f(\bm{X})\big)+\eta_{2}(\bm{X})\phi\big(-f(\bm{X})\big)\Big).
\end{equation*}

Now we introduce the generic conditional $\phi$-risk,
$$
Q_{\eta_{1},\eta_{2}}(\alpha)=\eta_{1}\phi(\alpha)+\eta_{2}\phi(-\alpha),
$$
where $\eta_{1},\eta_{2}\in\bR_+$ and $\alpha\in\bR$. The notation suppresses the dependence on $\phi$ and $g$. We define the optimal conditional $\phi$-risk,
$$
H(\eta_1,\eta_2) = Q_{\eta_1,\eta_2}(\alpha^{\ast})=\min_{\alpha\in\bR}Q_{\eta_1,\eta_2}(\alpha),
$$
and furthermore define,
$$
H^-(\eta_1,\eta_2) = \min_{\alpha:\alpha(\eta_1-\eta_2)\leq0}Q_{\eta_1,\eta_2}(\alpha).
$$
$H^-(\eta_1,\eta_2)$ is the optimal value of the conditional $\phi$-risk, under the constraint that the sign of the argument $\alpha$ disagrees with the Bayes regime.
Fisher consistency is equivalent to $H^-(\eta_1,\eta_2)>H(\eta_1,\eta_2)$ for any $\eta_1, \eta_2\in[0,\infty)$ with $\eta_1\neq\eta_2$.
The condition is similar with that of classification calibration in \cite{Bartlett2006:convexity}. When $\phi$ is convex, this condition is equivalent to a simpler condition on the derivative of $\phi$ at 0.
\begin{theorem}
\label{thm:fisher}
Assume that $\phi$ is convex. Then $\phi$ is Fisher consistent if and only if $\phi'(0)$ exists and $\phi'(0)<0$.
\end{theorem}
It is interesting to note that the necessary and sufficient condition for a convex surrogate loss function $\phi$ to yield a Fisher consistent regime concerns only its local property at $0$. All surrogate loss function listed above are Fisher consistent.

\subsubsection{Relating excess risk to excess $\phi$-risk}
We now turn to the excess risk and show how it can be bounded through the excess $\phi$-risk. It is easy to verify that the excess $\phi$-risk can be expressed as
$$
\Delta\Rcal_{\phi,g}(f) = \bE\left(Q_{\eta_1(\bm{X}),\eta_2(\bm{X})}(f(\bm{X})) - \min_{\alpha\in\bR}Q_{\eta_1(\bm{X}),\eta_2(\bm{X})}(\alpha)\right).
$$
Let  $\Delta Q_{\eta_1,\eta_2}(f) = Q_{\eta_1,\eta_2}(f)-\min_{\alpha\in\bR}Q_{\eta_1,\eta_2}(\alpha)=Q_{\eta_1,\eta_2}(f)-H_{\eta_1,\eta_2}.$

\begin{theorem}\label{thm:excessrisk}
Assume $\phi$ is convex, $\phi'(0)$ exists and $\phi'(0) < 0$. In addition, suppose that there exist constants $C>0$ and $s\ge 1$ such that
\begin{equation*}
|\eta_1-\eta_2|^s\leq C^s\Delta Q_{\eta_1,\eta_2}(0),
\end{equation*}
Then
$$
\Delta\Rcal(f)\leq C\left(\Delta\Rcal_{\phi,g}(f)\right)^{1/s}.
$$
\end{theorem}

As shown in the examples below, $\Delta Q_{\eta_1,\eta_2}(0)$ is often related to $\eta_1+\eta_2$. The following theorem handles this situation.
\begin{theorem}\label{thm:excessrisk2}
Assume $\phi$ is convex, $\phi'(0)$ exists, and $\phi'(0) < 0$. Suppose $\bE\left(\frac{|{R}_g|}{\pi(A,\bm{X})}\right)\leq M_g$. In addition, suppose that there exist a constant $s\ge 2$ and a concave increasing function $h: \bR_+\rightarrow\bR_+$ such that
\begin{equation*}
{|\eta_1-\eta_2|^s}\leq h(\eta_1+\eta_2)\Delta Q_{\eta_1,\eta_2}(0),
\end{equation*}
Then
$$
\Delta\Rcal(f)\leq \left(h(M_g)\right)^{1/s}\left(\Delta\Rcal_{\phi,g}(f)\right)^{1/s}.
$$
\end{theorem}

We now examine the consequences of these theorems on the examples of loss functions. Here we only present results briefly, and show details in Appendix A. Except for Examples 1 and 6, we assume that $\bE\left(\frac{|{R}_g|}{\pi(A,\bm{X})}\right)$ is bounded by $M_g$ in all other examples.

\textit{Example 1} (hinge loss). As shown in Appendix A, $H_{\eta_1,\eta_2} = 2\min(\eta_1,\eta_2)$, and $\Delta Q_{\eta_1,\eta_2}(0) = |\eta_1-\eta_2|$. By Theorem \ref{thm:excessrisk}, $\Delta\Rcal(f)\leq\Delta\Rcal_{\phi,g}(f)$.

\textit{Example 2} (squared hinge loss).  Consider the loss function $\phi(\alpha)=[(1-\alpha)_+]^2$. We have $(\eta_1-\eta_2)^2=(\eta_1+\eta_2)\Delta Q_{\eta_1,\eta_2}(0)$. By Theorem \ref{thm:excessrisk2}, $\Delta\Rcal(f)\leq\sqrt{M_g}(\Delta\Rcal_{\phi,g}(f))^{1/2}$.

\textit{Example 3} (least squares loss).  Now consider the loss function $\phi(\alpha)=(1-\alpha)^2$. Both $H_{\eta_1,\eta_2}$ and $\Delta Q_{\eta_1,\eta_2}(0)$ are the same as those in the previous example. Hence the bound in the previous example also applies to the least squares loss.

\textit{Example 4} (Huberized hinge loss). We can simply obtain that $(\eta_1-\eta_2)^2=4(\eta_1+\eta_2)\Delta Q_{\eta_1,\eta_2}(0)$. By Theorem \ref{thm:excessrisk2},
\begin{equation} \label{eq:hhlossbound}
\Delta\Rcal(f)\leq2\sqrt{M_g}(\Delta\Rcal_{\phi,g}(f))^{1/2}.
\end{equation}

\textit{Example 5} (logistic loss).  We consider the loss function $\phi(\alpha)=\log(1+\exp(-\alpha))$. This is a little complicated case. As shown in Appendix A, $(\eta_1-\eta_2)^2 \leq 8(\eta_1+\eta_2)\Delta Q_{\eta_1,\eta_2}(0)$. Then by Theorem \ref{thm:excessrisk2}, we have $\Delta\Rcal(f)\leq\sqrt{8M_g}(\Delta\Rcal_{\phi,g}(f))^{1/2}$.

\textit{Example 6} (DWD loss). As shown in Appendix A, we obtain $\Delta Q_{\eta_1,\eta_2}(0)\geq|\eta_1-\eta_2|$. Then by Theorem \ref{thm:excessrisk}, $\Delta\Rcal(f)\leq\Delta\Rcal_{\phi,g}(f)$.

\textit{Example 7} (exponential loss). Consider the loss function $\phi(\alpha)=\exp(-\alpha)$. We have $H_{\eta_1,\eta_2}=2\sqrt{\eta_1\eta_2}$, and $\Delta Q_{\eta_1,\eta_2}(0)=(\sqrt{\eta_1}-\sqrt{\eta_2})^2$. Then $(\eta_1-\eta_2)^2 \leq 2(\eta_1+\eta_2)\Delta Q_{\eta_1,\eta_2}(0)$. By Theorem \ref{thm:excessrisk2}, $\Delta\Rcal(f)\leq\sqrt{2M_g}(\Delta\Rcal_{\phi,g}(f))^{1/2}$.

\subsubsection{Universal consistency}
We will establish universal consistency of the regime $d_{f_{D_n,\lambda_n}}=\textrm{sign}(f_{D_n,\lambda_n})$. The following theorem shows the convergence of $\phi$-risk on the sample dependent function $f_{D_n,\lambda_n}$. We apply empirical process techniques to show consistency.
\begin{theorem}\label{thm:consistphirisk}
Suppose $\phi$ is a Lipschitz continuous function. Assume that we choose a sequence $\lambda_n>0$ such that $\lambda_n\rightarrow0$ and $n\lambda_n\rightarrow\infty$. For any distribution $P$ for $(\bm{X},A,R)$ satisfying $\frac{|{R}_g|}{\pi(A,\bm{X})}\leq M_g < \infty$ and $|\sqrt{\lambda_n}b_{D_n,\lambda_n}|\leq M_b< \infty$ almost everywhere, we have that in probability,
\begin{equation*}
\lim_{n\rightarrow\infty}\Rcal_{\phi,g}(f_{D_n,\lambda_n})=\inf_{f\in\Hcal_K+\{1\}}\Rcal_{\phi,g}(f).
\end{equation*}
\end{theorem}

When the loss function $\phi$ satisfies Theorem \ref{thm:excessrisk} or \ref{thm:excessrisk2}, starting from Theorem \ref{thm:consistphirisk}, universally consistent follows if $\inf_{f\in\Hcal_K+\{1\}}\Rcal_{\phi,g}(f)=\Rcal_{\phi,g}^{\ast}$. This condition requires the concept of universal kernels \citep{Steinwart:SVM2008}.
A continuous kernel $K$ on a compact metric space $\Xcal$ is called universal if its associated RKHS $\Hcal_K$ is dense in $C(\Xcal)$, the space of all continuous functions $f:\Xcal\rightarrow\bR$ on the compact metric space $\Xcal$ endowed with the usual supremum norm. 
The next Lemma shows that the RKHS $\Hcal_K$ of a universal kernel $K$ is rich enough to approximate arbitrary decision functions.
\begin{lemma}\label{thm:infphirisk}
Let $K$ be a universal kernel, and $\Hcal_K$ be the associated RKHS.
Suppose that $\phi$ is a Lipschitz continuous function, and $f_{\phi,g}^{\ast}$ is measurable and bounded, $|f_{\phi,g}^{\ast}|\leq M_f$.
For any distribution $P$ for $(\bm{X},A,R)$ satisfying $\frac{|{R}_g|}{\pi(A,\bm{X})}\leq M_g < \infty$ almost everywhere with regular marginal distribution on $\bm{X}$, we have
\begin{equation*}
\inf_{f\in\Hcal_K+\{1\}}\Rcal_{\phi,g}(f) = \Rcal_{\phi,g}^{\ast}.
\end{equation*}
\end{lemma}

Our proposed AOL uses the Huberized hinge loss.
Combining all the theoretical results and the excess risk bound in \eqref{eq:hhlossbound} together, the following proposition shows universal consistency of AOL with the Huberized hinge loss.
\begin{proposition}\label{thm:logisticconsist}
Let $K$ be a universal kernel, and $\Hcal_K$ be the associated RKHS. Let $\phi$ be the Huberized hinge loss function.
Assume that we choose a sequence $\lambda_n>0$ such that $\lambda_n\rightarrow0$ and $n\lambda_n\rightarrow\infty$. For any distribution $P$ for $(\bm{X},A,R)$ satisfying $\frac{|{R}_g|}{\pi(A,\bm{X})}\leq M_g < \infty$ almost everywhere with regular marginal distribution on $\bm{X}$, we have that in probability,
\begin{equation*}
\lim_{n\rightarrow\infty}\Rcal(\textrm{sign}(f_{D_n,\lambda_n}))=\Rcal^{\ast}.
\end{equation*}
\end{proposition}
In the proof of Proposition \ref{thm:logisticconsist}, we provide a bound on $b_{D_n,\lambda_n}$. The similar trick can be applied to hinge loss, squared hinge loss, and least squares loss. Thus for these three loss functions, it is not hard to derive their universal consistency. The exponential loss function is not Lipschitz continuous, so the learning regime with this loss is probably not universally consistent.

For the logistic loss and DWD loss, they do not satisfy Lemma \ref{thm:infphirisk} since $f_{\phi,g}^{\ast}$ is not bounded. We require stronger conditions for consistency. Firstly, we may assume that both $\eta_1(\bm{x})$ and $\eta_2(\bm{x})$ in (\ref{eq:eta}) are continuous. This assumption is plausible in practice. Secondly, we still need an assumption on bounded $b_{D_n,\lambda_n}$ as in Theorem \ref{thm:consistphirisk} to exclude some trivial situations, for example, where $A\cdot\textrm{sign}({R}_g) = 1$ almost everywhere. We present the result in the following proposition. The proof is simple and we omit it.
\begin{proposition}\label{thm:logisticconsist}
Let $K$ be a universal kernel, and $\Hcal_K$ be the associated RKHS. Let $\phi$ be the logistic loss or the DWD loss.
Assume that we choose a sequence $\lambda_n>0$ such that $\lambda_n\rightarrow0$ and $n\lambda_n\rightarrow\infty$. For any distribution $P$ for $(\bm{X},A,R)$ satisfying that (1) both $\eta_1(\bm{x})$ and $\eta_2(\bm{x})$ are continuous, (2) $|\sqrt{\lambda_n}b_{D_n,\lambda_n}|\leq M_b< \infty$ almost everywhere, and (3) $\frac{|{R}_g|}{\pi(A,\bm{X})}\leq M_g < \infty$ almost everywhere with regular marginal distribution on $\bm{X}$, we have that in probability,
\begin{equation*}
\lim_{n\rightarrow\infty}\Rcal(\textrm{sign}(f_{D_n,\lambda_n}))=\Rcal^{\ast}.
\end{equation*}
\end{proposition}

\subsection{Variable selection for AOL} \label{sec:AOLvs}

As demonstrated in \citet{Zhou2015:RWL}, variable selection is critical for optimal treatment regime when the dimension of clinical covariates is moderate or high. In this section, we apply the variable selection techniques in \citet{Zhou2015:RWL} to AOL.

\subsubsection{Variable selection for linear AOL}
As in \citet{Zhou2015:RWL}, we apply the elastic-net penalty \citep{Zou2005:elasticnet},
\begin{equation*}
\lambda_1||\bm{w}||_1 + \frac{\lambda_2}{2}\bm{w}^T\bm{w},
\end{equation*}
where $||\bm{w}||_1=\sum_{j=1}^p|w_j|$ is the $\ell_1$-norm, to replace the $\ell_2$-norm penalty in (\ref{qieq:linearAOL}) for variable selection. The elastic-net penalty selects informative covariates through
the $\ell_1$-norm penalty, and tends to identify or remove highly correlated variables together, the so-called grouping property, as the $\ell_2$-norm penalty does.

The elastic-net penalized linear AOL minimizes
\begin{equation*}
\frac{1}{n}\sum_{i=1}^{n}\frac{|\tilde{r}_i|}{\pi(a_i,\bm{x}_i)}\phi\Big(a_i\cdot\textrm{sign}(\tilde{r}_i)\big(\bm{w}^T\bm{x}_i+b\big)\Big)+
\lambda_1||\bm{w}||_1 + \frac{\lambda_2}{2}\bm{w}^T\bm{w},
\end{equation*}
where $\lambda_1(>0)$ and $\lambda_2(\geq0)$ are regularization parameters.
We use projected scaled sub-gradient (PSS) algorithms \citep{Schmidt2010:LASSO}, which are extensions of L-BFGS to the case of optimizing a smooth function with an $\ell_1$-norm penalty.
The obtained decision function is $\hat{f}(\bm{x})=\hat{\bm{w}}^T\bm{x}+\hat{b}$, and thus the estimated optimal treatment regime is the sign of $\hat{f}(x)$.



\subsubsection{Variable selection for AOL with nonlinear kernels}
Similar in \citet{Zhou2015:RWL}, taking the Gaussian RBF kernel as an example, we define the covariates-scaled Gaussian RBF kernel,
\begin{equation*}
K_{\bm{\eta}}(\bm{x},\bm{z}) = \exp\left(-\sum_{j=1}^{p}\eta_j(x_j-z_j)^2\right),
\end{equation*}
where $\bm{\eta}=(\eta_1,\cdots,\eta_p)^T\geq\bm{0}$. The covariate $x_j$ is scaled by $\sqrt{\eta_j}$. Setting ${\eta_j}=0$ is equivalent to discarding the $j$'th covariate. The hyperparameter $\sigma$ in the original Gaussian RBF kernel is discarded as it is absorbed to the scaling factors. We seek $(\hat{\bm{v}}, \hat{b}, \hat{\bm{\eta}})$ to minimize the following optimization problem:
\begin{eqnarray} \label{nonlinearpenalizedAOLprime}
\min_{\bm{v},b,\bm{\eta}} &&\frac{1}{n}\sum_{i=1}^{n}\frac{|\tilde{r}_i|}{\pi(a_i,\bm{x}_i)}\phi\Big(a_i\cdot\textrm{sign}(\tilde{r}_i)\big(\sum_{j=1}^nv_jK_{\bm{\eta}}(\bm{x}_i,\bm{x}_j)+b\big)\Big) \nonumber \\
&&\quad + \lambda_1||\bm{\eta}||_1+ \frac{\lambda_2}{2}\sum_{i,j=1}^nv_iv_jK_{\bm{\eta}}(\bm{x}_i,\bm{x}_j),\\
\textrm{subject to} && \bm{\eta} \geq \bm{0}, \nonumber
\end{eqnarray}
where $\lambda_1(>0)$ and $\lambda_2(>0)$ are regularization parameters. There are $n+p+1$ variables for the optimization problem.
It has an $\ell_1$-norm penalty on scaling factors. It could yield zero solutions for some of the $\bm{\eta}$ due to the singularity at $\bm{\eta}=\bm{0}$, and hence performs variable selection. Note that the optimization problem (\ref{nonlinearpenalizedAOLprime}) is not convex any more, even if the loss function is convex.
We apply L-BFGS-B algorithm \citep{Byrd1995:LBFGSB, Morales2011:LBFGSB}, an extension of L-BFGS to handle simple box constraints on variables, to solve (\ref{nonlinearpenalizedAOLprime}).
Then the obtained decision function is $\hat{f}(\bm{x})=\sum_{i=1}^n\hat{v}_iK_{\hat{\bm{\eta}}}(\bm{x},\bm{x}_i)+\hat{b}$.

\section{Simulation studies}
We carried out extensive simulations to investigate empirical performance of the proposed AOL methods.

We first evaluated performance of different residuals in the framework of AOL.
In the simulations, $p$-dimensional vectors of clinical covariates $x_1,\cdots,x_p$ were generated from independent uniform random variables $U(-1,1)$.
The response $R$ was normally distributed with mean $Q_0({x},a)$ and standard deviation 1. We considered two scenarios with linear treatment regimes:
\begin{enumerate}
\item[(1)] $Q_0({x},a)=(0.5+0.5x_1+0.8x_2+0.3x_3-0.5x_4+0.7x_5) + a(0.2-0.6x_1-0.8x_2)$;
\item[(2)] $Q_0({x},a)=\exp\left[(0.5+0.5x_1+0.8x_2+0.3x_3-0.5x_4+0.7x_5) + a(0.2-0.6x_1-0.8x_2)\right]$.
\end{enumerate}
We evaluated three types of residuals, as discussed in Section \ref{sec:connection}, with respect to the following $g(\bm{x})$'s:
\begin{itemize}
\item $\tilde{g}(\bm{x})=\mathbb{E}\left(\frac{\pi(-A,\bm{X})}{\pi(A,\bm{X})}R\big|\bm{X}=\bm{x}\right) = \pi(-1,\bm{x})\mu_{+1}(\bm{x})+\pi(+1,\bm{x})\mu_{-1}(\bm{x})$;
\item ${g}_1(\bm{x}) = \mathbb{E}\left(\frac{R}{2\pi(A,X)}\big|\bm{X}=\bm{x}\right) = \frac{1}{2}\mu_{+1}(\bm{x})+\frac{1}{2}\mu_{-1}(\bm{x})$;
\item ${g}_2(\bm{x}) = \mathbb{E}\left(R|\bm{X}=\bm{x}\right) = \pi(+1,\bm{x})\mu_{+1}(\bm{x})+\pi(-1,\bm{x})\mu_{-1}(\bm{x})$.
\end{itemize}
$\tilde{g}(\bm{x})$ is used in the proposed AOL. However, $g_1(\bm{x})$ in \citet{Zhou2015:RWL} and $g_2(\bm{x})$ in \citet{Liu2014:ABSOWL} also can be applied in AOL to replace $\tilde{g}(\bm{x})$. In a randomized clinical trial with usual equal allocation ratio 1:1, \textit{i.e.} $\pi(+1,\bm{x})=\pi(-1,\bm{x})=0.5$, these $g(\bm{x})$'s are the same. To compare performance of these residuals, we considered unequal allocation ratios (1) 3:1, \textit{i.e.}, $\pi(+1,\bm{x})=3\pi(-1,\bm{x})$ and (2) 1:3, \textit{i.e.}, $3\pi(+1,\bm{x})=\pi(-1,\bm{x})$.

The sample sizes were $n=100$ and $n=400$ for each scenario. We repeated the simulation 500 times. Parameters were tuned through 10-fold cross-validation. A large independent test set with 10,000 subjects was used to evaluate performance. The evaluation criterion was the value function of the estimated regime on the test set.

For simplicity, we run the first set of simulations using only linear AOL on low dimensional data ($p=5$). $\tilde{g}(\bm{x})$, ${g}_1(\bm{x})$ and ${g}_2(\bm{x})$ are obtained by the underlying true distributions of the data, instead of estimating them from the observed data, to eliminate impacts of regression estimates on evaluation.

The simulation results on low dimensional data ($p=5$) are presented in Table~\ref{tab:residuals}. From the table, the residuals from $\tilde{g}(\bm{x})$ yielded the best performance for each combination of scenario, allocation ratio and sample size, especially when the sample size is small. The simulation results confirm finite sample performance of our proposed counterfactual residual.

\begin{table}[tbp]
\centering
\small
\caption{Mean (std) of empirical value functions evaluated on independent test data for AOL with three types of residuals in two simulation scenarios with 5 covariates. The best value function for each scenario and sample size combination is in bold.}
\label{tab:residuals}
\begin{tabular}{@{}lccccc@{}}
\addlinespace
\toprule
& \multicolumn{2}{c}{allocation ratio = 3:1} & \phantom{a} & \multicolumn{2}{c}{allocation ratio = 1:3} \\
& {$n=100$} & {$n=400$} & \phantom{a} & {$n=100$} & {$n=400$} \\
\midrule
& \multicolumn{5}{c}{Scenario 1 (Optimal value $1.001$)} \\
\midrule
$\tilde{g}(\bm{x})$ & \textbf{0.940 (0.076)} & \textbf{0.986 (0.018)} && \textbf{0.954 (0.042)} & \textbf{0.974 (0.023)} \\
${g}_1(\bm{x})$ & 0.927 (0.087) & 0.985 (0.022) && 0.945 (0.050) & \textbf{0.974 (0.022)} \\
${g}_2(\bm{x})$ & 0.894 (0.101) & 0.978 (0.026) && 0.919 (0.066) & 0.970 (0.022) \\
\midrule
& \multicolumn{5}{c}{Scenario 2 (Optimal value $3.659$)} \\
\midrule
$\tilde{g}(\bm{x})$ & \textbf{3.375 (0.331)} & \textbf{3.451 (0.269)} && \textbf{3.626 (0.034)} & \textbf{3.655 (0.013)} \\
${g}_1(\bm{x})$ & 3.363 (0.336) & 3.447 (0.282) && 3.615 (0.052) & 3.650 (0.015)  \\
${g}_2(\bm{x})$ & 3.314 (0.347) & 3.430 (0.281) && 3.584 (0.072) & 3.631 (0.021) \\
\bottomrule
\end{tabular}
\end{table}

We then compared performance of AOL with other existing methods on usual equal allocation ratio data. The treatment $A\in\mathcal{A}=\{-1, 1\}$ was independent of ${X}$ with $\pi(+1,\bm{X})=\pi(-1,\bm{X})=0.5$. The covariate $\bm{X}$ and the outcome $R$ were generated as previously. We considered two additional scenarios with non-linear treatment regimes:
\begin{enumerate}
\item[(3)] $Q_0({x},a)=(0.5+0.6x_1+0.8x_2+0.3x_3-0.5x_4+0.7x_5) + a(0.6-x_1^2-x_2^2)$;
\item[(4)] $Q_0({x},a)=\exp\left[(0.5+0.6x_1+0.8x_2+0.3x_3-0.5x_4+0.7x_5) + a(0.6-x_1^2-x_2^2)\right]$;
\end{enumerate}

We run simulations for two different dimensions of covariates: low dimensional data ($p=5$) and moderate dimensional data ($p=25$). On low dimensional data ($p=5$), we compared empirical performances of the following seven methods: (1) $\ell_1$-PLS proposed by \citet{Qian:ITR2011}; (2) Q-learning using random forests as described in \citet{Taylor2015:QRF} (Q-RF);
(3) Doubly robust augmented inverse probability weighted estimator (AIPWE) with CART proposed by \citet{Zhang2012:ClassificationITR} (AIPWE-CART);
(4) RWL proposed in \citet{Zhou2015:RWL} using the linear kernel (RWL-Linear); (5) RWL using the Gaussian RBF kernel (RWL-Gaussian);
(6) the proposed AOL using the linear kernel (AOL-Linear); (7) the proposed AOL using the Gaussian RBF kernel (AOL-Gaussian).
When the dimension was moderate ($p=25$), RWL methods were replaced with their variable selection counterparts (RWL-VS-Linear and RWL-VS-Gaussian) \citep{Zhou2015:RWL}, and similarly AOL methods were replaced with AOL-VS-Linear and AOL-VS-Gaussian.

$\ell_1$-PLS is a parametric regression-based method. In the simulation studies, $\ell_1$-PLS estimated the conditional outcomes $\mathbb{E}(R|\bm{X},A)$ by a linear model on $(1,\bm{X},A,\bm{X}A)$, and used the LASSO penalty for variable selection. The obtained regime was the treatment arm in which the conditional mean outcome is larger.
Q-RF is a nonparametric regression-based method. The conditional outcomes $\mathbb{E}(R|\bm{X},A)$ were approximated using $(\bm{X},A)$ as input covariates in the random forests. The number of trees was set to 1000 as suggested in \citet{Taylor2015:QRF}.
For AIPWE-CART, we first obtained the AIPWE version of the contrast function through linear regression, and then we let the propensity score be 0.5 and searched the optimal treatment regime using a CART. The residuals in RWL and AOL were the same, and they were estimated by a linear regression model on $\bm{X}$. It was different with the previous simulation. We pretended that we do not know the underlying distribution of the data, and the residuals were estimated purely based on the simulated data. There are tuning parameters for $\ell_1$-PLS, RWL and AOL methods. Parameters were tuned through 10-fold cross-validation.

\begin{table}[tbp]
\centering
\small
\caption{Mean (std) of empirical value functions evaluated on independent test data for 4 simulation scenarios with 5 covariates. The best value function for each scenario and sample size combination is in bold.}
\label{tab:value5}
\begin{tabular}{@{}lccccc@{}}
\addlinespace
\toprule
& {$n=100$} & {$n=400$} & \phantom{a} & {$n=100$} & {$n=400$} \\
\midrule
& \multicolumn{2}{c}{Scenario 1} & \phantom{a} & \multicolumn{2}{c}{Scenario 2} \\
& \multicolumn{2}{c}{(Optimal value $1.001$)} & \phantom{a} & \multicolumn{2}{c}{(Optimal value $3.659$)} \\
\midrule
$\ell_1$-PLS & \textbf{0.974 (0.021)} & \textbf{0.993 (0.006)} && 3.537 (0.069) & 3.549 (0.041) \\
Q-RF & 0.889 (0.053) & 0.952 (0.014) && 3.459 (0.137) & 3.588 (0.023) \\
AIPWE-CART & 0.855 (0.078) & 0.917 (0.034) && 3.307 (0.211) & 3.503 (0.061) \\
RWL-Linear & 0.930 (0.068) & 0.978 (0.018) && \textbf{3.565 (0.109)} & \textbf{3.640 (0.027)} \\
RWL-Gaussian & 0.909 (0.077) & 0.973 (0.023) && 3.516 (0.126) & 3.621 (0.042) \\
AOL-Linear & 0.946 (0.051) & 0.985 (0.014) && 3.546 (0.125) & 3.620 (0.030) \\
AOL-Gaussian & 0.907 (0.082) & 0.977 (0.023) && 3.517 (0.121) & 3.621 (0.037) \\
\midrule
& \multicolumn{2}{c}{Scenario 3} & \phantom{a} & \multicolumn{2}{c}{Scenario 4} \\
& \multicolumn{2}{c}{(Optimal value $0.848$)} & \phantom{a} & \multicolumn{2}{c}{(Optimal value $3.237$)} \\
\midrule
Q-RF & 0.619 (0.053) & 0.730 (0.028) &&  2.898 (0.151) & 3.127 (0.039)    \\
AIPWE-CART & 0.620 (0.086) & 0.740 (0.047) && {2.909 (0.171)} & 3.118 (0.056) \\
RWL-Gaussian &  0.638 (0.070) & 0.763 (0.041) &&  2.894 (0.130) & 3.125 (0.061) \\
AOL-Gaussian & \textbf{0.650 (0.070)} & \textbf{0.784 (0.041)} && \textbf{2.918 (0.135)} & \textbf{3.152 (0.054)} \\
\bottomrule
\end{tabular}
\end{table}

Again, the sample sizes were $n=100$ and $n=400$ for each scenario. We repeated the simulation 500 times.  A large independent test set with 10,000 subjects was used to evaluate performance.

The simulation results on the low dimensional data ($p=5$) are presented in Table~\ref{tab:value5}.
For Scenario 1, the optimal regime $d^{\ast}(\bm{x})$ is $1$ if $0.6x_1+0.8x_2<0.2$, and $-1$ otherwise. 
Both the decision boundary and the conditional outcome were linear. Thus $\ell_1$-PLS performed very well since its model was correctly specified. RWL and AOL methods performed similarly, and they were close to $\ell_1$-PLS especially when the sample size was large. Q-RF and AIPWE-CART, as tree-based methods, were not comparable with other methods, perhaps trees do not work well to detect linear boundary.
For Scenario 2, the optimal treatment regime was the same as the one in Scenario 1. Although the boundary was linear, both the conditional outcome and the contrast function were non-linear. $\ell_1$-PLS did not yield the best performance due to model mis-specification. Instead RWL-Linear showed the best performance.  AOL-Linear was slightly worse than RWL-Linear. We used a linear model to estimate residuals. In this scenario, the linear model for residuals was mis-specified. RWL is robust to outliers on the residuals \citep{Zhou2015:RWL}. As discussed in Section \ref{sec:connection}, the robustness to outliers is the only nice property that AOL does not inherit from RWL due to unbounded convex loss function.
This is perhaps the reason why AOL-Linear was slightly worse than RWL.
Both RWL-Gaussian and AOL-Gaussian were similarly performed.
For Scenarios 3 and 4 , the decision boundaries were both nonlinear. We show results for Q-RF, AIPWE-CART, RWL-Gaussian, and AOL-Gaussian since the other three methods, $\ell_1$-PLS, RWL-Linear and AOL-Linear, can only detect linear regimes. In Scenario 3, the model to estimating residuals was correctly specified, while in Scenario 4, this model was mis-specified. For both scenarios, AOL-Gaussian yielded the best performance, and was slightly better than RWL-Gaussian. The non-convex optimization with RWL-Gaussian has a complicated objective function with many local minima or stationary points. It is very challenging to find a global minimum. The convex AOL-Gaussian does not have such problem, and hence received better performance than RWL-Gaussian.
We also compared running times of RWL and AOL methods. As shown in Table~5 in Appendix C, AOL is about 5-10 times faster than RWL. The convex optimization is much more computationally efficient than non-convex optimization.

We moved to moderate dimension cases ($p=25$). The simulation results are shown in Table~\ref{tab:value25}. In Scenario 1, $\ell_1$-PLS outperformed other methods because of correct model specification. When the sample size was large, RWL and AOL methods were all close to $\ell_1$-PLS. In Scenario 2, RWL-VS-Linear presented the best performance, and were slightly better than AOL-VS-Linear. We think the reason is that RWL methods are robust on mis-specified regression models for estimating residuals. In Scenarios 3 and 4, our proposed AOL-VS-Gaussian ranked the first, and was slightly better than RWL-VS-Gaussian.
Even though both RWL-VS-Gaussian and AOL-VS-Gaussian involve non-convex optimization, the objective function of AOL-VS-Gaussian is simpler, and is perhaps easier to find a global minimum than RWL-VS-Gaussian. We also compared the computational costs of RWL and AOL, as shown in Table~6 in Appendix C. The cost of AOL was again about 5-10 times cheaper than that of RWL.

\begin{table}[tbp]
\centering
\small
\caption{Mean (std) of empirical value functions evaluated on independent test data for 4 simulation scenarios with 25 covariates. The best value function for each scenario and sample size combination is in bold.}
\label{tab:value25}
\begin{tabular}{@{}lccccc@{}}
\addlinespace
\toprule
& {$n=100$} & {$n=400$} & \phantom{a} & {$n=100$} & {$n=400$} \\
\midrule
& \multicolumn{2}{c}{Scenario 1} & \phantom{a} & \multicolumn{2}{c}{Scenario 2} \\
& \multicolumn{2}{c}{(Optimal value $1.001$)} & \phantom{a} & \multicolumn{2}{c}{(Optimal value $3.659$)} \\
\midrule
$\ell_1$-PLS &  \textbf{0.960 (0.036)} & \textbf{0.992 (0.007)} && 3.423 (0.057) & 3.531 (0.036) \\
Q-RF & 0.785 (0.081) & 0.926 (0.027) && 3.070 (0.375) & 3.529 (0.047) \\
AIPWE-CART &  0.794 (0.108) & 0.904 (0.038) && 3.307 (0.211) & 3.503 (0.061)\\
RWL-VS-Linear &  0.869 (0.085) & 0.973 (0.020) && \textbf{3.450 (0.177)} & \textbf{3.632 (0.033)}\\
RWL-VS-Gaussian & 0.846 (0.110) & 0.963 (0.038) && 3.399 (0.234) & 3.611 (0.049) \\
AOL-VS-Linear & 0.878 (0.082) & 0.976 (0.018) && 3.434 (0.153) & 3.591 (0.044) \\
AOL-VS-Gaussian & 0.861 (0.106) & 0.975 (0.039) && 3.421 (0.209) & 3.616 (0.037) \\
\midrule
& \multicolumn{2}{c}{Scenario 3} & \phantom{a} & \multicolumn{2}{c}{Scenario 4} \\
& \multicolumn{2}{c}{(Optimal value $0.848$)} & \phantom{a} & \multicolumn{2}{c}{(Optimal value $3.237$)} \\
\midrule
Q-RF & 0.541 (0.041) & 0.646 (0.038) && 2.678 (0.185) & 3.006 (0.074)  \\
AIPWE-CART & 0.542 (0.062) & 0.705 (0.068) && 2.724 (0.198) & 3.094 (0.066) \\
RWL-VS-Gaussian  & 0.559 (0.067) & 0.766 (0.061)   && 2.716 (0.205) & 3.166 (0.057) \\
AOL-VS-Gaussian & \textbf{0.560 (0.067)} & \textbf{0.774 (0.070)}   && \textbf{2.735 (0.209)} & \textbf{3.168 (0.078)}  \\
\bottomrule
\end{tabular}
\end{table}

\section{Data analysis}
We applied the proposed methods to analyze the Nefazodone-CBASP clinical trial data \citep{Keller:depression}. The Nefazodone-CBASP trial randomly assigned patients with non-psychotic chronic major depressive disorder (MDD) in a 1:1:1 allocation ratio to either Nefazodone (NFZ), cognitive behavioral-analysis system of psychotherapy (CBASP), or the combination of Nefazodone and CBASP (COMB). The outcome was the score on the 24-item Hamilton Rating Scale for Depression (HRSD). Lower HRSD is better. We used 50 pre-treatment covariates as in \citet{Zhao:OWL2012}, and excluded patients with missing covariate values. The data used here consisted of 647 patients, with 216, 220, and 211 patients in three treatment arms.

We performed pairwise comparisons between any two treatment arms.
we compared the performance of AOL-VS-Linear and AOL-VS-Gaussian with $l_1$-PLS, Q-RF, AIPWE-CART, RWL-VS-Linear and RWL-VS-Gaussian, as in the simulation studies.
The outcomes used in the analyses were opposites of HRSD scores.
We used a nested 10-fold cross-validation procedure for an unbiased comparison \citep{Ambroise:bias}. Specifically, the data were randomly partitioned into 10 roughly equal-sized parts. We used nine parts as training data to predict optimal treatments for patients in the part left out. The parameter tuning was based on inner 10-fold cross-validation on the training data. We repeated the procedure 10 times, and obtained the predicted treatment for each patient. We then computed the estimated value function as $\mathbb{P}_n[R\mathbb{I}(A=Pred)/\pi_A(\bm{X})]/\mathbb{P}_n[\mathbb{I}(A=Pred)/\pi_A(\bm{X})]$, where $\mathbb{P}_n$ denotes the empirical average over the data and $Pred$ is the predicted treatment in the cross validation procedure.
To obtain reliable estimates, we repeated the nested cross-validation procedure 100 times with different fold partitions.

The analysis results are presented in Table~\ref{tab:depression}. For comparison between NFZ and CBASP, RWL-VS-Linear and AOL-VS-Linear performed better than other methods. For comparisons between NFZ and COMB and between CBASP and COMB, all methods produced similar performance. AOL-VS-Linear was among the top two methods for all comparisons. As shown in Table 7 in Appendix C, AOL was at least 10 times faster than RWL.

\begin{table}[tbp]
\centering
\caption{Mean score (standard deviation) on HRSD from the cross-validation procedure using different methods. Lower HRSD score is better. The two best scores for each comparison is in bold.}%
\begin{tabular}{@{}lcccc@{}}
\addlinespace
\toprule
& NFZ vs CBASP & NFZ vs COMB & CBASP vs COMB \\
\midrule
$\ell_1$-PLS & 16.30 (0.39) & 11.20 (0.16) & 10.95 (0.09) \\
Q-RF &  16.27 (0.44) & 11.05 (0.18) & 10.93 (0.09) \\
AIPWE & 16.45 (0.41) & \textbf{10.97 (0.15)} & 10.96 (0.14) \\
RWL-VS-Linear & \textbf{15.45 (0.37)} & 11.09 (0.29) & \textbf{10.88 (0.05)} \\
RWL-VS-Gaussian & 16.29 (0.44) & 11.33 (0.25) & 11.07 (0.28) \\
AOL-VS-Linear & \textbf{15.77 (0.37)} & \textbf{11.03 (0.18)} & \textbf{10.90 (0.06)} \\
AOL-VS-Gaussian & 16.32 (0.36) & 11.21 (0.23) & 11.02 (0.16) \\
\bottomrule
\end{tabular}
\label{tab:depression}
\end{table}

\section{Discussion}
In this article, we have proposed augmented outcome-weighted learning (AOL) to estimate optimal treatment regimes. As a close relative of residual weighted learning (RWL), AOL possesses almost all nice properties of RWL. AOL utilizes a convex loss function to guarantee a global solution. By contrast, the nice theoretical properties, for example, universal consistency, of RWL rely on a global solution, but the non-convex optimization associated with RWL cannot guarantee the global optimization.
Furthermore, the convex optimization associated with AOL make it computationally efficient. In the simulation studies and data analysis, AOL is at least 5-10 times faster than RWL.

There are two main approaches to estimating optimal treatment regimes. Regression-based approaches posit regression models for either conditional mean outcomes, $\mu_{+1}(\bm{x})$ and $\mu_{-1}(\bm{x})$, or the contrast function, $\delta(\bm{x})=\mu_{+1}(\bm{x})-\mu_{-1}(\bm{x})$, then the optimal treatment regime is estimated by setting $\delta(\bm{x})=0$.
Classification-based approaches directly estimate the optimal regime $\delta(\bm{x})=0$ in a semiparametric or nonparametric model. Compared with regression-based approaches, classification-based approaches are more robust to model misspecification. For example, in Figure~\ref{fig:example}, the optimal treatment regime $\delta(\bm{x})=0$ is almost linear, but the contrast function $\delta(\bm{x})$ is a complicated non-linear function. We may use the linear AOL to estimate the regime. However, any regression-based approach with a linear model would be misspecified. Another example is Scenario 2 in the simulation studies, where the optimal regime is linear, but neither the conditional mean outcome nor the contrast function is linear. The linear AOL yielded better performance than $\ell_1$-PLS, although both methods posit a linear model.


AOL uses a different form of residuals, as compared to RWL. The residuals are estimated with respect to a counterfactual average outcome where all subjects would receive the opposite treatments to what they have actually received. Apparently, both AOL and RWL can apply with any form of residuals. In this article, we focus on the randomized clinical trial data where $\pi(a,\bm{x})$ is known. According to the theory in Section \ref{sec:theory}, for a randomized clinical trial, AOL with any form of residuals is universally consistent. That is, it would eventually yield the Bayes regime when the sample size approaches infinity. When the sample size is finite, the simulations in Section 3 confirm the better performance of the counterfactual residual over others. The counterfactual residual is derived from a doubly robust AIPWE. The double robustness is quite useful for the observational study. In Appendix D, we develop the double robustness on universal consistency, \textit{i.e.},  the estimated regime of AOL is universal consistent if either $\hat{\mu}_a(\bm{x})$ or $\hat{\pi}(a,\bm{x})$ is consistent on the observational data. We pave the way in theory for AOL in observational studies. It is of great interest to apply AOL in observational studies in our future work.

In this article, the outcome $R$ is continuous. \citet{Zhou2015:RWL} proposed RWL as a general framework to deal with continuous, binary, count and rate outcomes. Similarly, AOL can handle all these types of outcomes by calculating residuals from a weighted regression model. The only difference is that the weights of AOL are $\frac{\pi(-A,\bm{X})}{\pi(A,\bm{X})}$, while $\frac{1}{2\pi(A,\bm{X})}$ for RWL.

Variable selection is critical for optimal treatment regimes \citep{Zhou2015:RWL}. Similar with RWL, we have provided variable selection algorithms for both linear and Gaussian RBF kernels. The variable selection with the linear kernel is an important extension of AOL. It seeks the optimal treatment regime semiparametrically, which is suitable for high dimensional data. Unlike the linear kernel, the variable selection with the Gaussian RBF kernel involves computationally intensive non-convex optimization, which cannot guarantee a global solution. A convex extension is still needed for our future investigation. Perhaps we may apply with the Gaussian RBF kernel a similar adaptive metric selection in \citet{Zhou2017:KNN}.

\appendix
\makeatletter   
 \renewcommand{\@seccntformat}[1]{APPENDIX~{\csname the#1\endcsname}.\hspace*{1em}}
 \makeatother

\vspace{40pt}
\noindent \textbf{\LARGE APPENDIX}

\vspace{10pt}
\noindent We investigate loss functions in Appendix A. The proofs of theorems in the main paper are given in Appendix B. We present additional simulation results in Appendix C. The doubly robustness of AOL on observational data is proved in Appendix D.

\section{Loss functions}
\textit{Example 1} (hinge loss). Consider the loss function $\phi(\alpha)=(1-\alpha)_+$. This is the surrogate loss function used in \cite{Zhao:OWL2012} and \cite{Liu2014:ABSOWL}. $Q_{\eta_1,\eta_2}(\alpha)$ is piecewise-linear. For $\eta_1=0$, any $\alpha\leq-1$ makes $Q_{\eta_1,\eta_2}(\alpha)$ vanish. The same holds for $\alpha\geq1$ for $\eta_2=0$. For $\eta_1,\eta_2\in(0,\infty)$, any minima lie in $[-1,1]$. Since $Q_{\eta_1,\eta_2}(\alpha)$ is linear on $[-1,1]$, the minimum must be attained at 1 for $\eta_1>\eta_2$, $-1$ for $\eta_1<\eta_2$, and anywhere in $[-1,1]$ for $\eta_1=\eta_2$. We have argued that $\alpha^{\ast} = \sign(\eta_1-\eta_2)$. It is easy to verify that $H(\eta_1,\eta_2)=2\min(\eta_1,\eta_2)$. Similar argument gives $H^-(\eta_1,\eta_2) = \eta_1+\eta_2$. $H^-(\eta_1,\eta_2)$ is strictly greater than $H(\eta_1,\eta_2)$ when $\eta_1\neq\eta_2$, so the hinge loss is Fisher consistent.
Since $H_{\eta_1,\eta_2} = 2\min(\eta_1,\eta_2)$, we have $\Delta Q_{\eta_1,\eta_2}(0) = |\eta_1-\eta_2|$. From Theorem \ref{thm:excessrisk}, $\Delta\Rcal(f)\leq\Delta\Rcal_{\phi,g}(f)$.

\textit{Example 2} (squared hinge loss).  Consider the loss function $\phi(\alpha)=[(1-\alpha)_+]^2$. This function is convex, differentiable, and decreasing at zero, and thus is Fisher consistent. If $\eta_1=0$, any $\alpha\leq-1$ makes $Q_{\eta_1,\eta_2}(\alpha)$ vanish. Similarly, any $\alpha\geq1$ makes the conditional $\phi$-risk vanish when $\eta_2=0$. For $\eta_1,\eta_2\in(0,\infty)$, $Q_{\eta_1,\eta_2}(\alpha)$ is strictly convex with a unique minimum, and solving for it yields $\alpha^{\ast}=(\eta_1-\eta_2)/(\eta_1+\eta_2)$. Simple calculation gives that $H_{\eta_1,\eta_2}$ is 0 when $\eta_1=\eta_2=0$, and otherwise $4\eta_1\eta_2/(\eta_1+\eta_2)$. Then for either case, we have $(\eta_1-\eta_2)^2=(\eta_1+\eta_2)\Delta Q_{\eta_1,\eta_2}(0)$.
If we further assume that $\bE\left(\frac{|{R}_g|}{\pi(A,\bm{X})}\right)$ is bounded by $M_g$, by Theorem \ref{thm:excessrisk2},
\begin{equation*}
\Delta\Rcal(f)\leq\sqrt{M_g}(\Delta\Rcal_{\phi,g}(f))^{1/2}.
\end{equation*}

\textit{Example 3} (least squares loss).  Now consider the loss function $\phi(\alpha)=(1-\alpha)^2$. From Theorem \ref{thm:fisher}, it is Fisher consistent. Simple algebraic manipulations show that $H_{\eta_1,\eta_2}$ and $\Delta Q_{\eta_1,\eta_2}(0)$ are the same as those in the previous example. Hence the bound in the previous example also applies to the least squares loss.

\textit{Example 4} (Huberized hinge loss). If $\eta_1=0$, any $\alpha\leq-1$ makes $Q_{\eta_1,\eta_2}(\alpha)$ vanish. Similarly, when $\eta_2=0$ any $\alpha\geq1$ makes the conditional $\phi$-risk vanish. For $\eta_1>0$ and $\eta_2 > 0$, $Q_{\eta_1,\eta_2}(\alpha)$ is strictly convex with a unique minimum. Solving by differentiation, the minimum is obtained at $\alpha^{\ast}=(\eta_1-\eta_2)/(\eta_1+\eta_2)$. Then we have $H_{\eta_1,\eta_2}$ is 0 when $\eta_1=\eta_2=0$, and otherwise $\eta_1\eta_2/(\eta_1+\eta_2)$. Then for either case, we have $(\eta_1-\eta_2)^2=4(\eta_1+\eta_2)\Delta Q_{\eta_1,\eta_2}(0)$.
If we further assume that $\bE\left(\frac{|{R}_g|}{\pi(A,\bm{X})}\right)$ is bounded by $M_g$, by Theorem \ref{thm:excessrisk2},
\begin{equation*}
\Delta\Rcal(f)\leq2\sqrt{M_g}(\Delta\Rcal_{\phi,g}(f))^{1/2}.
\end{equation*}

\textit{Example 5} (logistic loss).  We consider the loss function $\phi(\alpha)=\log(1+\exp(-\alpha))$. This loss function is convex, differentiable, and decreasing at zero, and thus is Fisher consistent. We first consider the case that $\eta_1\neq0$ and $\eta_2\neq0$. Simple calculation gives that $Q_{\eta_1,\eta_2}(\alpha)$ attains its minimum at $\alpha^{\ast}=\log(\eta_1/\eta_2)$, and
$$
\Delta Q_{\eta_1,\eta_2}(0) = \eta_1\log\left(\frac{2\eta_1}{\eta_1+\eta_2}\right) + \eta_2\log\left(\frac{2\eta_2}{\eta_1+\eta_2}\right).
$$
We fix $\eta_2$, and see $\Delta Q_{\eta_1,\eta_2}(0)$ as a function of $\eta_1$. Using Taylor expansion around $\eta_1=\eta_2$, we have
$$
\Delta Q_{\eta_1,\eta_2}(0) = \frac{1}{2}\frac{\eta_2}{\tilde{\eta}_1(\tilde{\eta}_1+\eta_2)}(\eta_1-\eta_2)^2,
$$
where $\tilde{\eta}_1$ is between $\eta_1$ and $\eta_2$. Similarly, fix $\eta_1$, and again use Taylor expansion around $\eta_2=\eta_1$,
$$
\Delta Q_{\eta_1,\eta_2}(0) = \frac{1}{2}\frac{\eta_1}{\tilde{\eta}_2(\eta_1+\tilde{\eta}_2)}(\eta_1-\eta_2)^2,
$$
where $\tilde{\eta}_2$ is between $\eta_1$ and $\eta_2$.
By summing these two equations, we obtain
\begin{eqnarray*}
2\Delta Q_{\eta_1,\eta_2}(0) & = & \frac{1}{2}\frac{\eta_2}{\tilde{\eta}_1(\tilde{\eta}_1+\eta_2)}(\eta_1-\eta_2)^2+\frac{1}{2}\frac{\eta_1}{\tilde{\eta}_2(\eta_1+\tilde{\eta}_2)}(\eta_1-\eta_2)^2\\
& \geq & \frac{1}{2}\frac{\eta_2}{2(\eta_1+\eta_2)^2}(\eta_1-\eta_2)^2+\frac{1}{2}\frac{\eta_1}{2(\eta_1+\eta_2)^2}(\eta_1-\eta_2)^2 \\
& = & \frac{(\eta_1-\eta_2)^2}{4(\eta_1+\eta_2)}.
\end{eqnarray*}
So, we have
$$
(\eta_1-\eta_2)^2 \leq 8(\eta_1+\eta_2)\Delta Q_{\eta_1,\eta_2}(0).
$$
It is easy to verify that when $\eta_1=0$ or $\eta_2=0$, the above bound holds. If we further assume that $\bE\left(\frac{|{R}_g|}{\pi(A,\bm{X})}\right)$ is bounded by $M_g$, by Theorem \ref{thm:excessrisk2},
\begin{equation*}
\Delta\Rcal(f)\leq\sqrt{8M_g}(\Delta\Rcal_{\phi,g}(f))^{1/2}.
\end{equation*}

\textit{Example 6} (DWD loss). The DWD loss is convex, differentiable and decreasing at zero. Hence this loss function is Fisher consistent. When $\eta_1$, $\eta_2 \in (0,\infty)$, consider three cases, (1) $\eta_1>\eta_2$; (2) $\eta_2>\eta_1$; (3) $\eta_1=\eta_2$. Simple differentiation yields that the minimizer is
\begin{equation*}
\alpha^{\ast} =
\left\{
\begin{array}{ll}
\sqrt{\frac{\eta_1}{\eta_2}} & {\rm if}\: \eta_1>\eta_2>0, \\
\textrm{any point } \in[-1,1] & {\rm if}\: \eta_1=\eta_2>0, \\
-\sqrt{\frac{\eta_2}{\eta_1}} & {\rm if}\: \eta_2>\eta_1>0.
\end{array}
\right.
\end{equation*}
When $\eta_1>\eta_2>0$, we have $H_{\eta_1,\eta_2}=2\eta_2+2\sqrt{\eta_1\eta_2}$. Then,
\begin{equation} \label{eq:dwddeltaQ}
\Delta Q_{\eta_1,\eta_2}(0)=2\eta_1-2\sqrt{\eta_1\eta_2} = \frac{2\sqrt{\eta_1}}{\sqrt{\eta_1}+\sqrt{\eta_2}}(\eta_1-\eta_2)\geq|\eta_1-\eta_2|.
\end{equation}
Similarly, when $\eta_2>\eta_1>0$, (\ref{eq:dwddeltaQ}) holds. It is easy to verify that when $\eta_1=\eta_2$ or at least one of $\eta_1$ and $\eta_2$ is zero, the inequality (\ref{eq:dwddeltaQ}) holds too. By Theorem \ref{thm:excessrisk}, $\Delta\Rcal(f)\leq\Delta\Rcal_{\phi,g}(f)$.

\textit{Example 7} (exponential loss). Consider the loss function $\phi(\alpha)=\exp(-\alpha)$. Again, this function is convex, differentiable, and decreasing at zero, and thus is Fisher consistent. For $\eta_1$, $\eta_2\in(0,\infty)$, solving for the stationary point yields the unique minimizer $ \alpha^{\ast} = \argmin_{\alpha\in\bR}Q_{\eta_1,\eta_2}(\alpha) = \frac{1}{2}\log\left(\eta_1/\eta_2\right)$. Then $H_{\eta_1,\eta_2}=2\sqrt{\eta_1\eta_2}$, and $\Delta Q_{\eta_1,\eta_2}(0)=(\sqrt{\eta_1}-\sqrt{\eta_2})^2$. Note that $(\sqrt{\eta_1}+\sqrt{\eta_2})^2\leq 2(\eta_1+\eta_2)$, then we have,
$$
(\eta_1-\eta_2)^2 \leq 2(\eta_1+\eta_2)\Delta Q_{\eta_1,\eta_2}(0).
$$
It is easy to verify that when $\eta_1=0$ or $\eta_2=0$, the above inequality holds. With an additional assumption that $\bE\left(\frac{|{R}_g|}{\pi(A,\bm{X})}\right)$ is bounded by $M_g$, from Theorem \ref{thm:excessrisk2},
\begin{equation*}
\Delta\Rcal(f)\leq\sqrt{2M_g}(\Delta\Rcal_{\phi,g}(f))^{1/2}.
\end{equation*}

\section{Proofs}

\textbf{Proof of Theorem \ref{thm:fisher}}
\begin{proof}
Recall that $Q_{\eta_1,\eta_2}(\alpha)=\eta_1\phi(\alpha)+ \eta_2\phi(-\alpha)$. It is easy to check that $Q_{\eta_1,\eta_2}$ is convex.

We consider the {\bf `if'} part of the proof first. Suppose that $\phi$ is differentiable at 0 and has $\phi'(0)<0$. Assume without loss of generality that $\eta_1 > \eta_2$. We need to prove that $Q_{\eta_1,\eta_2}(\alpha)$ is not minimized by any $\alpha\in (-\infty,0]$, \textit{i.e.} $\alpha^{\ast}>0$. Because $\phi$ is convex, it follows that for any $h>0$
\begin{eqnarray*}
\phi(0)+h\phi'(0)& \le& \phi(h) \notag\\
\phi(0)-h\phi'(0)& \le& \phi(-h).
\end{eqnarray*}
Therefore, noting that $Q_{\eta_1,\eta_2}(0)=\phi(0)(\eta_1+\eta_2)$, it is derived that
\begin{eqnarray*}
Q_{\eta_1,\eta_2}(-h)-Q_{\eta_1,\eta_2}(0)&=&\eta_1\left(\phi(-h)-\phi(0)\right)+\eta_2\left(\phi(h)-\phi(0)\right)\\
&\ge& -(\eta_1-\eta_2)\phi'(0)h,
\end{eqnarray*}
that is, given $\phi'(0)<0$, for any $h>0$, $Q_{\eta_1,\eta_2}(-h)-Q_{\eta_1,\eta_2}(0)> 0$. Consequently, $\alpha^*\ge0$ because it is a minimum. To prove the strict inequality, note that given that $\phi$ is differentiable at zero, by definition, for any $\epsilon>0$ there exists a $\delta(\epsilon)>0$ such that
\begin{eqnarray*}
\delta^{-1}\left(\phi(\delta)-\phi(0)\right)&\le& \phi'(0)+\epsilon \\
\delta^{-1}\left(\phi(0)-\phi(-\delta)\right)&\ge& \phi'(0)-\epsilon.
\end{eqnarray*}
This implies that
\begin{eqnarray*}
Q_{\eta_1,\eta_2}(\delta)-Q_{\eta_1,\eta_2}(0)&=&\eta_1\left(\phi(\delta)-\phi(0)\right)+\eta_2\left(\phi(-\delta)-\phi(0)\right)\\
&\le& \eta_1\delta\left(\phi'(0)+\epsilon\right)+\eta_2\delta\left(\epsilon - \phi'(0)\right)\\ & = &\delta\left(\phi'(0)(\eta_1-\eta_2)+\epsilon(\eta_1+\eta_2)\right),
\end{eqnarray*}
thus, making $\epsilon$ small enough, $\phi'(0)(\eta_1-\eta_2)+\epsilon(\eta_1+\eta_2)<0$. It follows that $\alpha^{\ast}>0$.\\

We proceed now with the {\bf `only if'} part of the proof. Suppose $\phi$ is Fisher consistent. Note that if $\alpha^{\ast}$ minimizes $Q_{\eta_1,\eta_2}(\alpha)$, it follows that $Q_{\eta_1,\eta_2}(\alpha^{\ast})-Q_{\eta_1,\eta_2}(0)<0$ when $\eta_1 \neq \eta_2$. Note that
\begin{equation}\label{eq:onlyif_positive}
Q_{\eta_1,\eta_2}(\alpha^{\ast})-Q_{\eta_1,\eta_2}(0) = \eta_1(\phi(\alpha^{\ast})-\phi(0))+\eta_2(\phi(-\alpha^{\ast})-\phi(0))
\end{equation}

We need to prove that $\phi'(0) < 0$. Let $[a,b]$ be the subderivative of $\phi$ at zero. By definition, if $h > 0$,
\begin{eqnarray} \label{eq:subderivative}
\phi(h)-\phi(0) \ge bh \notag\\
\phi(-h) -\phi(0) \ge -ah.
\end{eqnarray}
First we are going to prove that $b<0$. Suppose by contradiction that $b\ge0$. If $\eta_1>\eta_2$ then $\alpha^{\ast} > 0$ from the definition of Fisher consistency. By (\ref{eq:subderivative}), $\phi(\alpha^{\ast})\ge\phi(0)$, and replacing in (\ref{eq:onlyif_positive}), it is necessary that $\phi(-\alpha^{\ast}) < \phi(0)$ in order to keep the optimality property of $\alpha^{\ast}$. By (\ref{eq:subderivative}) again, we have that $0<a\le b$. Consequently, by replacing (\ref{eq:subderivative}) with $h=\alpha^{\ast}$ into (\ref{eq:onlyif_positive}),
$$
Q_{\eta_1,\eta_2}(\alpha^{\ast})-Q_{\eta_1,\eta_2}(0)\ge \alpha^{\ast}\left(b(\eta_1-\eta_2)-\eta_2(a-b)\right)>0,
$$
which contradicts that $\alpha^{\ast}$ is the minimum. Therefore, it is concluded that $b<0$.

It remains to prove that $a=b$.  To do so, suppose by contradiction that $a<b<0$. This implies that it is possible to have a distribution such that $\eta_1>\eta_2$ and $\eta_1b>\eta_2a$,  and therefore $\alpha^{\ast}>0$. By replacing (\ref{eq:subderivative}) with $h=\alpha^{\ast}$ into (\ref{eq:onlyif_positive}) again,
$$
Q_{\eta_1,\eta_2}(\alpha^{\ast})-Q_{\eta_1,\eta_2}(0)\ge \alpha^{\ast}\left(\eta_1b-\eta_2a\right)>0,
$$
which contradicts the fact that $\alpha^{\ast}$ minimizes $Q_{\eta_1,\eta_2}$. It follows that $\phi$ is differentiable at zero and $\phi'(0)<0$.
\end{proof}

\textbf{Proof of Theorem \ref{thm:excessrisk}}
\begin{proof}
\begin{eqnarray*}
&&\bE\left(\frac{R}{\pi(A,\bm{X})}\bI(A\neq\sign(f(\bm{X})))\Big|\bm{X}\right)-\bE\left(\frac{R}{\pi(A,\bm{X})}\bI(A\neq d^{\ast}(\bm{X}))\Big|\bm{X}\right)\\
&=&\Big(\bE(R|\bm{X},A=1)-\bE(R|\bm{X},A=-1)\Big)\Big(\bI\big(d^{\ast}(\bm{X})=1\big)-\bI\big(\sign(f(\bm{X}))=1\big)\Big)\\
&\leq&\Big|\eta_1(\bm{X})-\eta_2(\bm{X})\Big|\bI\Big(\sign(f(\bm{X}))\big(\eta_1(\bm{X})-\eta_2(\bm{X})\big)<0\Big).
\end{eqnarray*}
Then taking expectation on both sides we have,
\begin{eqnarray*}
\Delta\Rcal(f)&\leq&\bE\Big(\big|\eta_1(\bm{X})-\eta_2(\bm{X})\big|\bI\big(\sign(f(\bm{X}))\big(\eta_1(\bm{X})-\eta_2(\bm{X})\big)<0\big)\Big)\\
&\leq &\left(\bE\Big(\big|\eta_1(\bm{X})-\eta_2(\bm{X})\big|^s\bI\big(\sign(f(\bm{X}))\big(\eta_1(\bm{X})-\eta_2(\bm{X})\big)<0\big)\Big)\right)^{1/s}\\
&\leq&C\left(\bE\Big(\Delta Q_{\eta_1(\bm{X}),\eta_2(\bm{X})}(0)\bI\big(\sign(f(\bm{X}))\big(\eta_1(\bm{X})-\eta_2(\bm{X})\big)<0\big)\Big)\right)^{1/s}.
\end{eqnarray*}
The second inequality follows from the Jensen's inequality. By the conditions of $\phi$, $\phi$ is Fisher consistent. Let $\alpha^{\ast}$ minimizes $Q_{\eta_1,\eta_2}(\alpha)$. When $\sign(f)\cdot(\eta_1-\eta_2)<0$, by the definition of Fisher consistent, 0 is between $f$ and $\alpha^{\ast}$. The convexity of $\phi$, and hence of $Q_{\eta_1,\eta_2}$, implies that
$$
Q_{\eta_1,\eta_2}(0)\leq \max(Q_{\eta_1,\eta_2}(f), Q_{\eta_1,\eta_2}(\alpha^{\ast}))=Q_{\eta_1,\eta_2}(f).
$$
So we have,
\begin{eqnarray*}
\Delta\Rcal(f)&\leq&C\left(\bE\Big(\Delta Q_{\eta_1(\bm{X}),\eta_2(\bm{X})}(f)\bI\big(\sign(f(\bm{X}))\big(\eta_1(\bm{X})-\eta_2(\bm{X})\big)<0\big)\Big)\right)^{1/s}\\
&\leq&C(\Delta\Rcal_{\phi,g}(f))^{1/s}.
\end{eqnarray*}
\end{proof}

\textbf{Proof of Theorem \ref{thm:excessrisk2}}
\begin{proof} Following the similar arguments in the proof of Theorem \ref{thm:excessrisk}, we have
\begin{eqnarray*}
\Delta\Rcal(f)&=&\bE\Big(\big|\eta_1(\bm{X})-\eta_2(\bm{X})\big|\bI\big(\sign(f(\bm{X}))\big(\eta_1(\bm{X})-\eta_2(\bm{X})\big)<0\big)\Big)\\
&\leq &\left(\bE\Big(\big|\eta_1(\bm{X})-\eta_2(\bm{X})\big|^{s/2}\bI\big(\sign(f(\bm{X}))\big(\eta_1(\bm{X})-\eta_2(\bm{X})\big)<0\big)\Big)\right)^{2/s}\\
&\leq&\left(\bE\Big(h(\eta_1(\bm{X})+\eta_2(\bm{X}))^{1/2}\Delta Q_{\eta_1(\bm{X}),\eta_2(\bm{X})}(0)^{1/2}\bI\big(\sign(f(\bm{X}))\big(\eta_1(\bm{X})-\eta_2(\bm{X})\big)<0\big)\Big)\right)^{2/s}\\
&\leq&\left(\bE\Big(h(\eta_1(\bm{X})+\eta_2(\bm{X}))^{1/2}\Delta Q_{\eta_1(\bm{X}),\eta_2(\bm{X})}(f(\bm{X}))^{1/2}\Big)\right)^{2/s}\\
&\leq&\left(\bE\Big(h(\eta_1(\bm{X})+\eta_2(\bm{X}))\Big)\bE\Big(\Delta Q_{\eta_1(\bm{X}),\eta_2(\bm{X})}(f(\bm{X}))\Big)\right)^{1/s}\\
&\leq&\Big(h(\bE(\eta_1(\bm{X})+\eta_2(\bm{X})))\Big)^{1/s}(\Delta\Rcal_{\phi,g}(f))^{1/s}. 
\end{eqnarray*}
The second and sixth inequalities follows from the Jensen's inequality, and the fifth follows Cauchy-Schwarz inequality.
Notice that
\begin{eqnarray*}
\bE\left(\frac{|{R}_g|}{\pi(A,\bm{X})}\Big|\bm{X}\right) &= & \bE(|{R}_g||\bm{X},A=1)+\bE(|{R}_g||\bm{X},A=-1)\\
&=&\bE({R}_g^++{R}_g^-|\bm{X},A=1)+\bE({R}_g^++{R}_g^-|\bm{X},A=-1)\\
&=&\eta_1(\bm{X})+\eta_2(\bm{X}).
\end{eqnarray*}
So $\bE\left(\frac{|{R}_g|}{\pi(A,\bm{X})}\right)=\bE(\eta_1(\bm{X})+\eta_2(\bm{X}))$. The desired result follows through the monotonicity of $h$.
\end{proof}

\textbf{Proof of Theorem \ref{thm:consistphirisk}}
\begin{proof}
Let $L(h,b)=|{R}_g|\phi(A\textrm{sign}({R}_g)(h(\bm{X})+b))/\pi(A,\bm{X})$.
For simplicity, we denote $f_{D_n,\lambda_n}$, $h_{D_n,\lambda_n}$ and $b_{D_n,\lambda_n}$ by $f_n$, $h_n$ and $b_n$, respectively.
By the definition of $h_{D_n,\lambda_n}$ and $b_{D_n,\lambda_n}$, we have, for any $h\in\mathcal{H}_K$ and $b\in\mathbb{R}$,
\begin{equation*}
\mathbb{P}_n(L(h_n,b_n))\leq\mathbb{P}_n(L(h_n,b_n))+\frac{\lambda_n}{2}||h_n||^2_K \leq \mathbb{P}_n(L(h,b))+\frac{\lambda_n}{2}||h||^2_K,
\end{equation*}
where $\mathbb{P}_n$ denotes the empirical measure of the observed data.
Then,
$\lim\sup_n\mathbb{P}_n(L(h_n,b_n))\leq\mathbb{P}(L(h,b))=\mathcal{R}_{\phi,g}(h+b)$ with probability 1. This implies
\begin{equation*}
\lim\sup_n\mathbb{P}_n(L(h_n,b_n))\leq\inf_{h\in\mathcal{H}_K, b\in\mathbb{R}}\mathcal{R}_{\phi,g}(h+b)\leq \mathbb{P}(L(h_n,b_n))
\end{equation*}
with probability 1. It suffices to show $\mathbb{P}_n(L(h_n,b_n))-\mathbb{P}(L(h_n,b_n))\rightarrow0$ in probability.

We have a bound for $|b_n|$, $|\sqrt{\lambda_n}b_n|\leq M_b$, as a condition. We next obtain a bound for $||h_n||_K$. Since $\mathbb{P}_n(L(h_n,b_n))+\lambda_n||h_n||^2_K/2\leq\mathbb{P}_n(L(h,b))+\lambda_n||h||^2_K/2$, for any $h\in \mathcal{H}_K$ and $b\in\mathbb{R}$, we can choose  $h=0$ and $b=0$ to obtain,
$\mathbb{P}_n(L(h_n,b_n))+\lambda_n||h_n||^2_K/2\leq\phi(0)\mathbb{P}_n(|{R}_g|/\pi(A,\bm{X}))$. We thus have,
\begin{equation*}
\lambda_n||h_n||^2_K \leq 2\phi(0)\mathbb{P}_n(|{R}_g|/\pi(A,\bm{X})) \leq 2\phi(0)M_g.
\end{equation*}
Let $M_h=\sqrt{2\phi(0)M_g}$. Then the $\mathcal{H}_K$ norm of $\sqrt{\lambda_n}h_n$ is bounded by $M_h$.

Note that the class $\{\sqrt{\lambda_n}h:||\sqrt{\lambda_n}h||_K\leq M_h\}$ is a Donsker class. So $\{\sqrt{\lambda_n}(h+b):||\sqrt{\lambda_n}h||_K\leq M_h, |\sqrt{\lambda_n}b|\leq M_b\}$ is also P-Donsker.

Consider a function $\phi_{\lambda_n}(u) = \sqrt{\lambda_n}\phi(u/\sqrt{\lambda_n})$. $\phi_{\lambda_n}(u)$ is a Lipschitz continuous function with the same Lipschitz constant as $\phi(u)$. Note that
\begin{equation*}
\sqrt{\lambda_n}L(h,b) = \frac{|{R}_g|}{\pi(A,\bm{X})}\phi_{\lambda_n}(A\sqrt{\lambda_n}\cdot\textrm{sign}({R}_g)(h(\bm{X})+b)).
\end{equation*}
Since $\phi_{\lambda_n}(u)$ is Lipschitz continuous and $\frac{|{R}_g|}{\pi(A,\bm{X})}$ is bounded, the class $\{\sqrt{\lambda_n}L(h,b):||\sqrt{\lambda_n}h||_K\leq M_h, |\sqrt{\lambda_n}b|\leq M_b\}$ is also P-Donsker. Therefore
\begin{equation*}
\sqrt{n\lambda_n}(\mathbb{P}_n-\mathbb{P})L(h_n, b_n)=O_p(1).
\end{equation*}
Consequently, from $n\lambda_n\rightarrow\infty$, $\mathbb{P}_n(L(h_n,b_n))-\mathbb{P}(L(h_n,b_n))\rightarrow 0$ in probability.
\end{proof}

\textbf{Proof of Lemma \ref{thm:infphirisk}}
\begin{proof}
Fix any $0<\epsilon<1$. Suppose $\phi$ is Lipschitz continuous with Lipschitz constant $C$.
Let $\mu$ be the marginal distribution of $\bm{X}$. Since $\mu$ is regular and $f_{\phi,g}^{\ast}$ is measurable, using Lusin's theorem in measure theory, we know that $f_{\phi,g}^{\ast}$ can be approximated by a continuous function $f'(\bm{x})\in C(\mathcal{X})$ such that $\mu(f'(\bm{x})\neq f_{\phi,g}^{\ast}(\bm{x}))\leq \frac{\epsilon}{4CM_fM_g}$. Since $f_{\phi,g}^{\ast}$ is between $[-M_f,M_f]$, we may limit $f'(\bm{x})\in[-M_f,M_f]$ (otherwise, truncate $f'(\bm{x})$ with upper bound $M_f$ and lower bound $-M_f$).
Thus
\begin{eqnarray*}
&&\mathbb{E}\left(\frac{|{R}_g|}{\pi(A,\bm{X})}\phi\Big(A\cdot\textrm{sign}({R}_g)f'(\bm{X})\Big)\Big|\bm{X}=\bm{x}\right)
-\mathbb{E}\left(\frac{|{R}_g|}{\pi(A,\bm{X})}\phi\Big(A\textrm{sign}({R}_g)f_{\phi,g}^{\ast}(\bm{X})\Big)\Big|\bm{X}=\bm{x}\right)\\
&=& \eta_1(\bm{x})\Big[\phi(f'(\bm{x}))-\phi(f_{\phi,g}^{\ast}(\bm{x}))\Big]
+\eta_2(\bm{x})\Big[\phi(-f'(\bm{x}))-\phi(-f_{\phi,g}^{\ast}(\bm{x}))\Big]\\
&\leq & C\big(\eta_1(\bm{x})+\eta_2(\bm{x})\big)|f'(\bm{x})-f_{\phi,g}^{\ast}(\bm{x})|.
\end{eqnarray*}
The last inequality is due to the fact that $\phi$ is Lipschitz continuous. Then, we have,
\begin{eqnarray*}
&&\mathcal{R}_{\phi,g}(f')-\mathcal{R}_{\phi,g}^{\ast} = |\mathcal{R}_{\phi,g}(f')-\mathcal{R}_{\phi,g}^{\ast}|\\
&\leq&C\int \big(\eta_1(\bm{x})+\eta_2(\bm{x})\big)|f'(\bm{x})-f_{\phi,g}^{\ast}(\bm{x})| \mu(d\bm{x})\\
&=&C\int\bE\Big(\frac{|{R}_g|}{\pi(A,\bm{X})}\Big|\bm{X}=\bm{x}\Big)|f'(\bm{x})-f_{\phi,g}^{\ast}(\bm{x})|\bI(f'(\bm{x})\neq f_{\phi,g}^{\ast}(\bm{x}))\mu(d\bm{x})
\end{eqnarray*}
Since $\frac{|{R}_g|}{\pi(A,\bm{X})}\leq M_g$ and both $f'(\bm{x})$ and $f_{\phi,g}^{\ast}(\bm{x})$ are between $[-M_f,M_f]$, we have
\begin{equation*}
\mathcal{R}_{\phi,g}(f')-\mathcal{R}_{\phi,g}^{\ast} < \epsilon/2.
\end{equation*}

Since $K$ is universal, there exists a function $f''\in\mathcal{H}_K$ such that $||f''-f'||_{\infty}<\frac{\epsilon}{2CM_g}$. Similarly,
\begin{eqnarray*}
&& |\mathcal{R}_{\phi,g}(f'')-\mathcal{R}_{\phi,g}(f')|\\
&\leq&C\int \big(\eta_1(\bm{x})+\eta_2(\bm{x})\big)|f''(\bm{x})-f'(\bm{x})| \mu(d\bm{x})\\
&=&C\int\bE\Big(\frac{|{R}_g|}{\pi(A,\bm{X})}\Big|\bm{X}=\bm{x}\Big)|f''(\bm{x})-f'(\bm{x})|\mu(d\bm{x})\\
&<& \epsilon/2.
\end{eqnarray*}
By combining the two inequalities, we have
\begin{equation*}
\mathcal{R}_{T,g}(f'')-\mathcal{R}_{\phi,g}^{\ast}<\epsilon.
\end{equation*}
Noting that $f''\in\mathcal{H}_K$ and letting $\epsilon\rightarrow0$, we obtain the desired result.
\end{proof}

\textbf{Proof of Proposition \ref{thm:logisticconsist}}
\begin{proof}
When $\frac{|{R}_g|}{\pi(A,\bm{X})}$ is bounded, the excess risk is bounded as argued in Section 2.4.2,
\begin{equation} \label{eq:excessriskbound}
\Delta\Rcal(f)\leq2\sqrt{M_g}(\Delta\Rcal_{\phi,g}(f))^{1/2}.
\end{equation}

Next, we obtain a bound for $b_{f_{D_n,\lambda_n}}$. We use the notations in the proof of Theorem \ref{thm:consistphirisk}. We claim that there is a solution $(h_n,b_n)$ such that $h_n(\bm{x}_i)+b_n\in[-1,1]$ for some $i$. Suppose that there is another solution $(h'_n,b'_n)$ such that $|h_n(\bm{x}_i)+b_n|>1$ for all $i$. Let $D_1=\{i:A_i\textrm{sign}({R}_{g,i})=1, h'_n(\bm{X}_i)+b'_n<-1\}$ and $D_2=\{i:A_i\textrm{sign}({R}_{g,i})=-1, h'_n(\bm{X}_i)+b'_n>1\}$. Denote
\begin{equation*}
\alpha_1 = \sum_{i\in D_1} \frac{|{R}_{g,i}|}{\pi(A_i,\bm{X}_i)}, \quad \textrm{and} \quad
\alpha_2 = \sum_{i\in D_2} \frac{|44{R}_{g,i}|}{\pi(A_i,\bm{X}_i)}.
\end{equation*}
We show that $\alpha_1=\alpha_2$. Otherwise, when $\alpha_1>\alpha_2$, let
\begin{equation*}
\delta = \min_{i:h'_n(\bm{X}_i)+b'_n<-1} |h'_n(\bm{X}_i)+b'_n|.
\end{equation*}
Then set $h_n=h'_n$ and $b_n=b'_n+(\delta-1)$. It is easy to check that $(h_n,b_n)$ is a better solution than $(h'_n,b'_n)$, which is contradicted with the fact that $(h'_n,b'_n)$ is a solution. Similarly, when $\alpha_1<\alpha_2$, let
\begin{equation} \label{eq:delta}
\delta = \min_{i:h'_n(\bm{X}_i)+b'_n>1} |h'_n(\bm{X}_i)+b'_n|.
\end{equation}
Then set $h_n=h'_n$ and $b_n=b'_n-(\delta-1)$. Thus $(h_n,b_n)$ is a better solution than $(h'_n,b'_n)$. It is a contradiction again. So we have $\alpha_1=\alpha_2$. However, when we set $\delta$ as in (\ref{eq:delta}), $h_n=h'_n$ and $b_n=b'_n-(\delta-1)$. $(h_n,b_n)$ is a solution and satisfies our claim. Now if a solution $(h_n,b_n)$ satisfies our claim for subject $i_0$, we then have,
\begin{equation*}
|b_n|\leq 1 + |h_n(\bm{X}_{i_0})|\leq 1+||h_n||_{\infty}.
\end{equation*}
Note that $||h||_{\infty}\leq C_K||h||_K$. We have,
\begin{equation*}
|\sqrt{\lambda_n}b_n|\leq \sqrt{\lambda_n} + C_K\sqrt{\lambda_n}||h_n||_K.
\end{equation*}
As in the proof of Theorem \ref{thm:consistphirisk}, $\sqrt{\lambda_n}||h_n||_K$ is bounded.
Since $\lambda_n\rightarrow0$, and $C_K$ is bounded, we have $|\sqrt{\lambda_n}b_n|$ is bounded too. So by Theorem \ref{thm:consistphirisk},
\begin{equation} \label{eq:consistphirisk}
\lim_{n\rightarrow\infty}\Rcal_{\phi,g}(f_{D_n,\lambda_n})=\inf_{f\in\Hcal_K+\{1\}}\Rcal_{\phi,g}(f).
\end{equation}

By the argument in Appendix A, the optimal function
\begin{equation*}
f^{\ast}_{\phi,g}(\bm{x}) = \left\{
\begin{array}{ll}
0 & \textrm{if } \eta_1(\bm{x})=\eta_2(\bm{x})=0,\\
\displaystyle \frac{\eta_1(\bm{x})-\eta_2(\bm{x})}{\eta_1(\bm{x})+\eta_2(\bm{x})} & \textrm{otherwise}.
\end{array}
\right.
\end{equation*}
Clearly, $f^{\ast}_{\phi,g}$ is measurable, and $|f^{\ast}_{\phi,g}(\bm{x})|\leq1$. By Lemma \ref{thm:infphirisk}
\begin{equation} \label{eq:excessphirisk0}
\inf_{f\in\Hcal_K+\{1\}}\Rcal_{\phi,g}(f) = \Rcal_{\phi,g}^{\ast}.
\end{equation}
Combining (\ref{eq:excessriskbound}), (\ref{eq:consistphirisk}) and (\ref{eq:excessphirisk0}), we have the desired result.
\end{proof}

\section{Additional results in the simulation study and data analysis}

In this section, we compared the computational cost of AOL with RWL. In the simulation studies, the average running times of the first 10 runs of AOL and RWL with tuned parameters listed in Table~\ref{tab:runningtime5} for low dimensional data ($p=5$) and listed in Table~\ref{tab:runningtime5} for moderate dimensional data ($p=25$). AOL is about 5-10 times faster than RWL.

\begin{table}[p]
\centering
\small
\caption{Running times (in seconds) of AOL and RWL for 4 simulation scenarios on 5-covariate data.}
\label{tab:runningtime5}
\begin{tabular}{@{}lccccc@{}}
\addlinespace
\toprule
& \multicolumn{2}{c}{$n=100$} &\phantom{a}& \multicolumn{2}{c}{$n=400$} \\
\cline{2-3}\cline{5-6} \\[-5pt]
& AOL & RWL & & AOL & RWL \\
\midrule
& \multicolumn{5}{c}{Linear kernel}\\
Scenario 1 & 0.010 & 0.046 && 0.012 & 0.068\\
Scenario 2 & 0.011 & 0.062 && 0.013 & 0.073 \\
\midrule
& \multicolumn{5}{c}{Gaussian kernel}\\
Scenario 1 & 0.040 & 0.479 && 0.123 & 0.731\\
Scenario 2 & 0.078 & 0.657 && 0.191 & 1.012 \\
Scenario 3 & 0.104 & 0.440 && 0.098 & 0.888\\
Scenario 4 & 0.099 & 0.432 && 0.410 & 3.557\\
\bottomrule
\end{tabular}
\end{table}

\begin{table}[p]
\centering
\small
\caption{Running times (in seconds) of AOL and RWL with variable selection for 4 simulation scenarios on 25-covariate data.}
\label{tab:runningtime25}
\begin{tabular}{@{}lccccc@{}}
\addlinespace
\toprule
& \multicolumn{2}{c}{$n=100$} &\phantom{a}& \multicolumn{2}{c}{$n=400$} \\
\cline{2-3}\cline{5-6} \\[-5pt]
& AOL & RWL & & AOL & RWL \\
\midrule
& \multicolumn{5}{c}{Linear kernel}\\
Scenario 1 & 0.006 & 0.042 && 0.010 & 0.055 \\
Scenario 2 & 0.008 & 0.073 && 0.012 & 0.111\\
\midrule
& \multicolumn{5}{c}{Gaussian kernel}\\
Scenario 1 & 0.298 & 1.169 && 5.660 & 35.651 \\
Scenario 2 & 0.422 & 2.045 && 5.561 & 23.265 \\
Scenario 3 & 0.752 & 2.544 && 9.487 & 24.264 \\
Scenario 4 & 0.365 & 1.896 && 11.831 & 50.391 \\
\bottomrule
\end{tabular}
\end{table}

For the real data analysis in Section 4, we averaged the running times in the 10 runs of the first fold partition of cross-validation with tuned parameters. AOL is at least 10 times faster than RWL.

\begin{table}[p]
\centering
\caption{Running times (in seconds) of AOL and RWL with variable selection on the Nefazodone-CBASP clinical trial data.}
\begin{tabular}{@{}lcccccccc@{}}
\addlinespace
\toprule
& \multicolumn{2}{c}{NFZ vs CBASP} &\phantom{a}& \multicolumn{2}{c}{NFZ vs COMB} &\phantom{a}& \multicolumn{2}{c}{CBASP vs COMB} \\
\cline{2-3}\cline{5-6}\cline{8-9} \\[-5pt]
& AOL & RWL & & AOL & RWL & & AOL & RWL \\
\midrule
Linear kernel & 0.008 & 0.638 && 0.011 & 0.373 && 0.007 & 0.084 \\
Gaussian kernel & 25.334 & 390.773 && 21.598 & 283.986 && 10.445 & 96.627 \\
\bottomrule
\end{tabular}
\label{tab:depressiontime}
\end{table}

\section{Double robustness of AOL on observational data}

In the main paper, AOL is mainly applied to randomized clinical trial data. This method also can be used on observational data.
In an observational study, we first estimate $\pi(a,\bm{x})$ by, for example, $\hat\pi(a,\bm{x})$. Then we need to estimate $\tilde{g}(\bm{x})$ by
\begin{equation*}
\hat{\tilde{g}}(\bm{x})=\hat\pi(-1,\bm{x})\hat\mu_{+1}(\bm{x})+\hat\pi(+1,\bm{x})\hat\mu_{-1}(\bm{x}),
\end{equation*}
where  $\hat\mu_{+1}(\bm{x})$ and $\hat\mu_{-1}(\bm{x})$ are estimators of $\mu_{+1}(\bm{x})$ and $\mu_{-1}(\bm{x})$, respectively. $\tilde{g}(\bm{x})$ also can be estimated by weighted regression with weights $\hat{\pi}(-a,\bm{x})/\hat{\pi}(a,\bm{x})$.

Here we suppress the dependence on observed data $\mathcal{D}_n$ for notations $\hat\pi(a,\bm{x})$, $\hat\mu_{+1}(\bm{x})$, and $\hat\mu_{-1}(\bm{x})$. Suppose that when $n$ approaches infinity, $\hat\pi(a,\bm{x})\overset{p}{\to} \tilde\pi(a,\bm{x})$, $\hat\mu_{+1}(\bm{x})\overset{p}{\to}\tilde\mu_{+1}(\bm{x})$, and $\hat\mu_{-1}(\bm{x})\overset{p}{\to}\tilde\mu_{-1}(\bm{x})$. When $\hat\mu_{+1}(\bm{x})$ and $\hat\mu_{-1}(\bm{x})$ are consistent, $\tilde\mu_{+1}(\bm{x})=\mu_{+1}(\bm{x})$ and $\tilde\mu_{-1}(\bm{x})=\mu_{-1}(\bm{x})$. When $\hat\pi(a,\bm{x})$ is consistent, $\hat\pi(a,\bm{x})=\pi(a,\bm{x})$.

Again, for finite sample observational data $\mathcal{D}_n$, let $f_{D_n,\lambda_n}\in\Hcal_K+\{1\}$, \textit{i.e.} $f_{D_n,\lambda_n}=h_{D_n,\lambda_n}+b_{D_n,\lambda_n}$, where $h_{D_n,\lambda_n}\in\Hcal_K$ and $b_{D_n,\lambda_n}\in\bR$, be a global minimizer of the following optimization problem:
\begin{equation} \label{eq:AOLHk}
\min_{f=h+b\in\Hcal_K+\{1\}} \quad \frac{1}{n}\sum_{i=1}^n\frac{|{R}_{\hat{\tilde{g}},i}|}{\hat\pi(A_i,\bm{X}_i)}{\phi\Big(A_i\cdot\sign({R}_{\hat{\tilde{g}},i})f(\bm{X}_i)\Big)}+\frac{\lambda_n}{2}||h||_K^2,
\end{equation}
where ${R}_{\hat{\tilde{g}},i}=R_i-\hat{\tilde{g}}(\bm{X}_i)$. Here we suppress $\phi$ and $g$ from the notations of $f_{D_n,\lambda_n}$, $h_{D_n,\lambda_n}$ and $b_{D_n,\lambda_n}$.

Note that $|r|\phi(\textrm{sign}(r)f)$ is continuous with respect to $r$. By the law of large numbers and the continuous mapping theorem,
\begin{equation*}
\frac{1}{n}\sum_{i=1}^n\frac{|{R}_{\hat{\tilde{g}},i}|}{\hat\pi(A_i,\bm{X}_i)}{\phi\Big(A_i\cdot\sign({R}_{\hat{\tilde{g}},i})f(\bm{X}_i)\Big)} \overset{p}{\to} \bE\left(\frac{|{R}_g|}{\tilde\pi(A,\bm{X})}{\phi\Big(A\cdot\sign({R}_g)f(\bm{X})\Big)}\right),
\end{equation*}
where $R_g=R-g(\bm{X})$ and $g(\bm{x})=\tilde\pi(-1,\bm{x})\tilde\mu_{+1}(\bm{x})+\tilde\pi(+1,\bm{x})\tilde\mu_{-1}(\bm{x})$ in this section.
As in the case of the randomized clinical trial in the main paper, we define the risk function of a treatment regime $d$ as
\begin{equation*}
\mathcal{R}(d) = \bE\left(\frac{R}{\pi(A,\bm{X})}\mathbb{I}\Big(A\neq d(\bm{X})\Big)\right).
\end{equation*}
For observational data, we define the surrogate $\phi$-risk function:
\begin{equation} \label{eq:obsphirisk}
\Rcal_{\phi,g}(f)=\bE\left(\frac{|{R}_g|}{\tilde\pi(A,\bm{X})}{\phi\Big(A\cdot\sign({R}_g)f(\bm{X})\Big)}\right).
\end{equation}
Similarly, the minimal $\phi$-risk as $\Rcal_{\phi,g}^{\ast}=\inf_f\Rcal_{\phi,g}(f)$ and $f_{\phi,g}^{\ast}=\arg\min_f\Rcal_{\phi,g}(f)$.

The purpose of this section is to investigate the universal consistency of the associated regime of $f_{D_n,\lambda_n}$ on observational data. We check the theoretical properties in Section \ref{sec:theory} for observational data in parallel.

First, let us investigate the conditions for $\mathcal{R}(\textrm{sign}(f_{\phi,g}^{\ast})) = \mathcal{R}(d^{\ast})$. Define
\begin{eqnarray} \label{eq:tidleeta}
\tilde\eta_{1}(\bm{x}) & = & \mathbb{E}({R}_g^+|\bm{X}=\bm{x},A=+1)\frac{\pi(+1,\bm{x})}{\tilde{\pi}(+1,\bm{x})} + \mathbb{E}({R}_g^-|\bm{X}=\bm{x},A=-1)\frac{\pi(-1,\bm{x})}{\tilde{\pi}(-1,\bm{x})}, \nonumber \\
\tilde\eta_{2}(\bm{x}) & = & \mathbb{E}({R}_g^+|\bm{X}=\bm{x},A=-1)\frac{\pi(-1,\bm{x})}{\tilde{\pi}(-1,\bm{x})} + \mathbb{E}({R}_g^-|\bm{X}=\bm{x},A=+1)\frac{\pi(+1,\bm{x})}{\tilde{\pi}(+1,\bm{x})}.
\end{eqnarray}
The $\phi$-risk can be expressed as
\begin{equation*}
\Rcal_{\phi,g}(f)=\bE\Big(\tilde\eta_{1}(\bm{X})\phi\big(f(\bm{X})\big)+\tilde\eta_{2}(\bm{X})\phi\big(-f(\bm{X})\big)\Big).
\end{equation*}
The condition in Theorem \ref{thm:fisher}, \textit{i.e.}, $\phi'(0)$ exists and $\phi'(0)<0$, only guarantees that $f_{\phi,g}^{\ast}(\bm{x})$ has the same sign as $\tilde\eta_{1}(\bm{x})-\tilde\eta_{2}(\bm{x})$.

When $\hat\pi(a,\bm{x})$ is consistent, it is obvious that $\tilde\eta_{1}(\bm{x})-\tilde\eta_{2}(\bm{x})=\mu_{+1}(\bm{x})-\mu_{-1}(\bm{x})$. When $\hat\mu_{+1}(\bm{x})$ and $\hat\mu_{-1}(\bm{x})$ are consistent, we have $g(\bm{x})=\tilde\pi(-1,\bm{x})\mu_{+1}(\bm{x})+\tilde\pi(+1,\bm{x})\mu_{-1}(\bm{x})$, and
\begin{equation*}
\tilde\eta_{1}(\bm{x})-\tilde\eta_{2}(\bm{x}) = [\mu_{+1}(\bm{x})-g(\bm{x})]\frac{\pi(+1,\bm{x})}{\tilde{\pi}(+1,\bm{x})}-[\mu_{-1}(\bm{x})-g(\bm{x})]\frac{\pi(-1,\bm{x})}{\tilde{\pi}(-1,\bm{x})}=\mu_{+1}(\bm{x})-\mu_{-1}(\bm{x}).
\end{equation*}
The conditions for $\mathcal{R}(\textrm{sign}(f_{\phi,g}^{\ast})) = \mathcal{R}(d^{\ast})$ are (i) either $\hat\mu_{a}(\bm{x})$ or $\hat\pi(a,\bm{x})$ is consistent; and (ii) $\phi'(0)$ exists and $\phi'(0)<0$.

For the excess risk bound, we can easily verify that Theorem \ref{thm:excessrisk} holds for observational data with an additional condition as follows.
\begin{theorem}\label{thm:obsexcessrisk}
Assume either $\hat\mu_{a}(\bm{x})$ or $\hat\pi(a,\bm{x})$ is consistent, $\phi$ is convex, $\phi'(0)$ exists and $\phi'(0) < 0$. In addition, suppose that there exist constants $C>0$ and $s\ge 1$ such that
\begin{equation*}
|\tilde\eta_1-\tilde\eta_2|^s\leq C^s\Delta Q_{\tilde\eta_1,\tilde\eta_2}(0),
\end{equation*}
Then
$$
\Delta\Rcal(f)\leq C\left(\Delta\Rcal_{\phi,g}(f)\right)^{1/s}.
$$
\end{theorem}

Theorem \ref{thm:excessrisk2} can be modified for observational data as follows by noticing that
\begin{equation*}
\tilde\eta_{1}(\bm{x})+\tilde\eta_{2}(\bm{x}) = \bE\left(\frac{|{R}_g|}{\tilde\pi(A,\bm{X})}\Big|\bm{X}=\bm{x}\right).
\end{equation*}
\begin{theorem}\label{thm:obsexcessrisk2}
Assume either $\hat\mu_{a}(\bm{x})$ or $\hat\pi(a,\bm{x})$ is consistent, $\phi$ is convex, $\phi'(0)$ exists, and $\phi'(0) < 0$. Suppose $\bE\left(\frac{|{R}_g|}{\tilde\pi(A,\bm{X})}\right)\leq M_g$. In addition, suppose that there exist a constant $s\ge 2$ and a concave increasing function $h: \bR_+\rightarrow\bR_+$ such that
\begin{equation*}
{|\tilde\eta_1-\tilde\eta_2|^s}\leq h(\tilde\eta_1+\tilde\eta_2)\Delta Q_{\tilde\eta_1,\tilde\eta_2}(0),
\end{equation*}
Then
$$
\Delta\Rcal(f)\leq \left(h(M_g)\right)^{1/s}\left(\Delta\Rcal_{\phi,g}(f)\right)^{1/s}.
$$
\end{theorem}
Thus, all excess risk bounds provided in Section 2.5.2 apply to observational data when either $\hat\mu_{a}(\bm{x})$ or $\hat\pi(a,\bm{x})$ is consistent.

For universal consistency, the following theorem is just the counterpart of Theorem \ref{thm:consistphirisk} on observational data.
\begin{theorem}\label{thm:obsconsistphirisk}
Suppose $\phi$ is a Lipschitz continuous function. Assume that we choose a sequence $\lambda_n>0$ such that $\lambda_n\rightarrow0$ and $n\lambda_n\rightarrow\infty$. For any distribution $P$ for $(\bm{X},A,R)$ satisfying $\frac{|{R}_{\hat{\tilde{g}}}|}{\hat\pi(A,\bm{X})}\leq M < \infty$ and $|\sqrt{\lambda_n}b_{D_n,\lambda_n}|\leq M_b< \infty$ almost everywhere, we have that in probability,
\begin{equation*}
\lim_{n\rightarrow\infty}\Rcal_{\phi,g}(f_{D_n,\lambda_n})=\inf_{f\in\Hcal_K+\{1\}}\Rcal_{\phi,g}(f).
\end{equation*}
\end{theorem}
The proof follows the idea in the proof of Theorem \ref{thm:consistphirisk}. We only show a sketch.
\begin{proof}
Let $L_n(h,b)=|{R}_{\hat{\tilde{g}}}|\phi(A\textrm{sign}({R}_{\hat{\tilde{g}}})(h(\bm{X})+b))/\hat\pi(A,\bm{X})$ and $L(h,b)=|{R}_g|\phi(A\textrm{sign}({R}_g)(h(\bm{X})+b))/\tilde\pi(A,\bm{X})$.
By the continuous mapping theorem, $L_n(h,b)\overset{p}{\to}L(h,b)$. So $\mathbb{P}L_n(h,b){\to}\mathbb{P}L(h,b)$.

For any $h\in\mathcal{H}_K$ and $b\in\mathbb{R}$, we have
\begin{equation*}
\mathbb{P}_n(L_n(h_n,b_n))\leq\mathbb{P}_n(L_n(h_n,b_n))+\frac{\lambda_n}{2}||h_n||^2_K \leq \mathbb{P}_n(L_n(h,b))+\frac{\lambda_n}{2}||h||^2_K.
\end{equation*}
Then, $\lim\sup_n\mathbb{P}_n(L(h_n,b_n))\leq\mathbb{P}(L(h,b))=\mathcal{R}_{\phi,g}(h+b)$ with probability 1. This implies
\begin{equation*}
\lim\sup_n\mathbb{P}_n(L_n(h_n,b_n))\leq\inf_{h\in\mathcal{H}_K, b\in\mathbb{R}}\mathcal{R}_{\phi,g}(h+b)\leq \mathbb{P}(L(h_n,b_n))
\end{equation*}
with probability 1. It suffices to show $\mathbb{P}_n(L_n(h_n,b_n))-\mathbb{P}(L_n(h_n,b_n))\rightarrow0$ in probability since $\mathbb{P}L_n(h,b){\to}\mathbb{P}L(h,b)$.

Similar as in the proof of Theorem \ref{thm:consistphirisk}, we derive a bound for $h_n$ as
\begin{equation*}
\lambda_n||h_n||^2_K \leq 2\phi(0)\mathbb{P}_n(|{R}_g|/\pi(A,\bm{X})) \leq 2\phi(0)M.
\end{equation*}
Let $M_h=\sqrt{2\phi(0)M}$. Then, $\{\sqrt{\lambda_n}(h+b):||\sqrt{\lambda_n}h||_K\leq M_h, |\sqrt{\lambda_n}b|\leq M_b\}$ is P-Donsker.

Consider a function $\phi_{\lambda_n}(u) = \sqrt{\lambda_n}\phi(u/\sqrt{\lambda_n})$. Note that
\begin{equation*}
\sqrt{\lambda_n}L_n(h,b) = \frac{|{R}_{\hat{\tilde{g}}}|}{\hat\pi(A,\bm{X})}\phi_{\lambda_n}(A\sqrt{\lambda_n}\cdot\textrm{sign}({R}_{\hat{\tilde{g}}})(h(\bm{X})+b)).
\end{equation*}
Since $\phi_{\lambda_n}(u)$ is Lipschitz continuous and $\frac{|{R}_{\hat{\tilde{g}}}|}{\hat\pi(A,\bm{X})}$ is bounded, the class $\{\sqrt{\lambda_n}L_n(h,b):||\sqrt{\lambda_n}h||_K\leq M_h, |\sqrt{\lambda_n}b|\leq M_b\}$ is also P-Donsker. Therefore
\begin{equation*}
\sqrt{n\lambda_n}(\mathbb{P}_n-\mathbb{P})L(h_n, b_n)=O_p(1).
\end{equation*}
Consequently, from $n\lambda_n\rightarrow\infty$, $\mathbb{P}_n(L(h_n,b_n))-\mathbb{P}(L(h_n,b_n))\rightarrow 0$ in probability.
\end{proof}

Following the similar arguments in Section 2.5.3, for the Huberized hinge loss, we have universal consistency of AOL on observational data.
\begin{proposition}\label{thm:obslogisticconsist}
Assume either $\hat\mu_{a}(\bm{x})$ or $\hat\pi(a,\bm{x})$ is consistent. Let $K$ be a universal kernel, and $\Hcal_K$ be the associated RKHS. Let $\phi$ be the Huberized hinge loss function.
Assume that we choose a sequence $\lambda_n>0$ such that $\lambda_n\rightarrow0$ and $n\lambda_n\rightarrow\infty$. For any distribution $P$ for $(\bm{X},A,R)$ satisfying $\frac{|{R}_{\hat{\tilde{g}}}|}{\hat\pi(A,\bm{X})}\leq M < \infty$ almost everywhere with regular marginal distribution on $\bm{X}$, we have that in probability,
\begin{equation*}
\lim_{n\rightarrow\infty}\Rcal(\textrm{sign}(f_{D_n,\lambda_n}))=\Rcal^{\ast}.
\end{equation*}
\end{proposition}
We may achieve universal consistency for other loss functions as we did in Section 2.5.3. AOL is doubly robust on universal consistency for observational data.

\bibliographystyle{jasa}
\bibliography{myreference}

\end{document}